\documentclass[12pt]{article}

\usepackage{epsfig}
\usepackage{lscape}

\usepackage{bm}

\newcommand{\eq}{\begin{equation}}
\newcommand{\eqx}{\end{equation}}
\newcommand{\eqn}{\begin{eqnarray}}
\newcommand{\eqnx}{\end{eqnarray}}
\newcommand{\nbm}{$\nBmax\ $}
\newcommand{\0}{\hphantom{0}}

\newcommand\txf[2]{{\textstyle{#1\over#2}}}
\newcommand\half{\txf12}
\newcommand\fourth{\txf14}
\newcommand\fh[1]{{#1\over2}}
\newcommand\Tr{\mathop{\rm Tr}\nolimits}

\newcommand\mod{\mathop{\rm mod}\nolimits}

\newcommand\bartildeg[1]{\tilde{\kern-.01em{\bar #1}}}
\newcommand\Phibartilde{\bartildeg{\Phi}}
\newcommand\Psibartilde{\bartildeg{\Psi}}
\newcommand\Thetabartilde{\bartildeg{\Theta}}
\newcommand\Omegabartilde{\bartildeg{\Omega}}
\newcommand\Xibartilde{\bartildeg{\Xi}}
\newcommand\bartilde[1]{\kern.2em{\tilde{\kern-.2em{\bar #1}}}}
\newcommand\Fbartilde{\bartilde{F}}
\newcommand\Abartilde{\bartilde{A}}
\newcommand\Bbartilde{\bartilde{B}}
\newcommand\Gbartilde{\bartilde{G}}
\newcommand\Ibartilde{\bartilde{I}}

\newcommand\threej[1]{
\left(\begin{array}{ccc}
#1
\end{array}\right)}
\newcommand\sixj[1]{
\left\{\begin{array}{ccc}
#1
\end{array}\right\}}
\newcommand\nBmax{n_{B\max}}

\chardef\TL=`\~ % tilde in a string
\chardef\UL=`\_ % underline character in a string

\begin{document}

\date{July 2, 2004.}

\title{
% \null\vskip -60 pt
\begin{flushright}
\normalsize\rm TPJU-9/04, IFUP-TH 2004/24
\end{flushright}
\vskip 10 pt
High precision study of the structure of $D=4$
supersymmetric Yang-Mills quantum mechanics}

\author{M. Campostrini\\
\normalsize \it INFN, Sezione di Pisa, and\\
\normalsize \it Dipartimento di Fisica ``Enrico Fermi''
                dell'Universit\`a di Pisa,\\
\normalsize \it Via Buonarroti 2, I-56127 Pisa, Italy\\
\normalsize and\\
J. Wosiek\\
\normalsize \it M.Smoluchowski Institute of Physics,
                Jagellonian University,\\
\normalsize \it Reymonta 4, 30-059 Krak\'{o}w, Poland}

\maketitle

\begin{abstract}
The spectrum of $D=4$ supersymmetric Yang-Mills quantum mechanics is
computed with high accuracy in all channels of angular momentum and
fermion number. Localized and non-localized states coexists in certain
channels as a consequence of the supersymmetric interactions with flat
valleys. All states fall into well identifiable supermultiplets
providing an explicit realization of supersymmetry on the
spectroscopic level. An accidental degeneracy among some
supermultiplets has been found.  Regularized Witten index converges to
a time-independent constant which agrees with earlier calculations.
\end{abstract}

% \vspace*{1cm}
% \noindent TPJU-7/04, IFUP-TH 2004/24
% \newline June 2004
% \newline hep-th/0407021

\newpage
\section{Introduction}

In the present paper we report detailed studies of the supersymmetric
Yang-Mills quantum mechanics (SYMQM) \cite{WI},\cite{CH}. The
particular model addressed here results from
the dimensional reduction of the supersymmetric Yang-Mills field
theory, with the SU(2) gauge group, from a four-dimensional space-time
($D=4$) to a single point in space. It is a member of a family of
quantum mechanical systems with the famous $D=10$,
SU$(N\rightarrow\infty)$, SYMQM at its upper end. The latter,
considered as a model of an M-theory \cite{BFSS}, attracted a lot of attention in
recent years, see \cite{BS} and \cite{WAT} for reviews and further references.
For that reason we have launched a
systematic, nonperturbative study of the whole family
(varying $D$ and $N$) in an attempt to understand their global
properties, and to develop adequate techniques while moving gradually
to more complex models \cite{JWNP}.  The detailed motivation and an
account of the relations to the M-theory can be found there.

One of the characteristic property of the supersymmetric quantum
mechanics with the Yang-Mills potential is appearance of the
continuous spectrum of non-localized states together with
discrete, localized bound states \cite{dWLN}.  The supersymmetric
vacuum is believed to be in the continuous sector, with discrete
spectrum beginning at some nonzero energy in general.
Interestingly, the $D=10$ (and not less than $10$) quantum
mechanics has also a threshold bound state at zero energy, which
agrees with the M-theory correspondence.  Apart of the $D=2$
\cite{CH},\cite{STS}, these systems are not soluble and the
overall picture just outlined has accumulated over the years of
intense studies of particular issues \cite{SM}-\cite{KAB3}.

In Ref.\ \cite{JWNP} we proposed to use the standard Hamiltonian
formulation of quantum mechanics. To this end we have constructed
explicitly the (finite) basis of gauge invariant states and
calculated algebraically matrix representations of a Hamiltonian
and other relevant observables (e.g., supersymmetry generators).
This done, we proceeded to compute numerically the complete
spectrum, the energy eigenstates, identified their supersymmetric
partners, computed Witten index, etc.  The method has an intrinsic
cutoff - the total number of allowed bosonic quanta \nbm . Since
our basis is the eigenbasis of the occupation number operators,
the cutoff is easy to implement. It is also gauge and rotationally
invariant, hence it preserves these important symmetries. Since
the size of the basis, i.e., the dimension of the cut Hilbert
space, grows rapidly with $n_B$, convergence with the cutoff is
the crucial question for this approach.  It turns out that in all
cases studied there (i.e., the Wess-Zumino quantum mechanics, and
$D=2$ and $D=4$, SU(2), SYMQM) many nontrivial results were
reliably obtained before the number of basis vectors grew out of
control \cite{KW}.  The approach applies as well to bosons and
fermions being entirely insensitive to the notorious sign problem
which plagues any Monte Carlo attempts to attack these systems.
Later on the new, recursive method of calculating matrix elements
significantly improved the precision of the solution of the $D=2$
SYMQM and eventually inspired the exact, analytic calculation of
the restricted Witten index for this model \cite{CW}.  To make
further progress one has to deal with the rapidly growing number
of states. Of course this problem is most severe in the $D=10$
model where some preliminary results for pure Yang-Mills system
were nevertheless already obtained confirming for example the
SO(9) invariance \cite{JW2}.

It this paper we have abandoned the brute force construction and
diagonalization of the Hamiltonian in the whole (cut) Hilbert
space. Instead, we have exploited fully the rotational invariance
solving the problem separately for each angular momentum. Second,
the recursive construction of matrix elements of Ref.\ \cite{CW}
was generalized and adapted to the fixed angular momentum channels
(Section III). The two tricks coupled together led to the
quantitative improvement of the precision and allowed to uncover a
rich structure of the system to a much more complete level
(Section IV).

Finally, for the scalar ($j=0$) sector, one can push the cutoff
even higher performing complete analytic separation of variables
in this case \cite{SAV,VABA}. Using the method adapted by van Baal
for the noncompact system considered here, one can reach yet
higher cutoffs in the two ($n_F=0,2$) channels. Results of this
approach will be briefly discussed in the next Section.

Recently, a new possibility to optimize numerical solutions for the lowest state
of the system has been investigated \cite{KAR}.

Effective Lagrangians for various dimensionally reduced
supersymmetric Yang-Mills theories, including SYMQM, have been
very recently derived in Ref.\ \cite{SM2}.

Supersymmetric Yang-Mills theories in extended space have been
studied for some time with the aid of the Hamiltonian approach on
the light cone \cite{HHPS}.

\section{The system and early results}

\subsection{Definitions}

The reduced quantum-mechanical Yang-Mills system is described by nine
canonically conjugate pairs of bosonic coordinates and momenta
$x^i_a(t)$, $p^i_a(t)$, $i=1,2,3$, $a=1,2,3$ and six independent
fermionic coordinates composing a Majorana spinor $\psi_a^\alpha(t)$,
$\alpha=1,...,4$, $a=1,2,3$ satisfying canonical anticommutation
relations.  In $D=4$, it is equally possible to impose the Weyl
condition and work with Weyl spinors. The Hamiltonian reads \cite{HS}
\begin{eqnarray}
H &=& H_K + H_P + H_F, \nonumber \\
H_K &=& {1\over 2} p_a^ip_a^i , \nonumber \\
H_P &=& {g^2\over 4}\epsilon_{abc}
\epsilon_{ade}x_b^i x_c^j x_d^i x_e^j, \label{eq:Hamiltonian} \\
\label{HD4} H_F &=& {i g \over 2}
\epsilon_{abc}\psi_a^T\Gamma^k\psi_b x_c^k; \nonumber
\end{eqnarray}
in $D=4$, $\Gamma^k$ are the standard Dirac $\alpha^k$ matrices. In all
explicit calculations we use the Majorana representation of
Ref.\ \cite{IZ}.

Even though three-dimensional space was reduced to a single point,
the system still has an internal Spin(3) rotational symmetry,
inherited from the original theory, and generated by the angular
momentum
\begin{equation}
J^i=\epsilon^{ijk}
\left(x^j_a p^k_a+{1\over 4}\psi^T_a\Sigma^{jk}\psi_a\right),
\label{JD4}
\end{equation}
with
\begin{equation}
\Sigma^{jk}=-{i\over 4}[\Gamma^j,\Gamma^k].
\end{equation}
Furthermore, the model posesses the residual of the local gauge
transformation generated by
\begin{equation}
G_a=\epsilon_{abc}\left(x_b^k p_c^k-{i\over 2}\psi^T_b\psi_c \right),
\label{GG}
\end{equation}
and it is invariant under the supersymmetry transformations with
the generators
\begin{equation}
Q_{\alpha}=(\Gamma^k\psi_a)_{\alpha}p^k_a + i g
\epsilon_{abc}(\Sigma^{jk}\psi_a)_{\alpha}x^j_b x^k_c.
\label{QD4}
\end{equation}
The bosonic potential $H_P$ in Eq.\ (\ref{eq:Hamiltonian}), when
written in a vector notation in the color space, has a form
\begin{equation}
V={g^2\over 4}\Sigma_{jk}(\vec{x}^j\times\vec{x}^k)^2, \label{eq:V}
\end{equation}
which exhibits the famous flat directions responsible for a rich
structure of the spectrum.

\subsection{Creation and annihilation operators}

The Hamiltonian (\ref{eq:Hamiltonian}) is polynomial in
momenta and coordinates. Therefore it is convenient to employ the
eigenbasis of the occupation number operators associated with all
degrees of freedom. To this end we rewrite bosonic and fermionic
variables in terms of creation and annihilation operators of simple,
normalized harmonic oscillators
\begin{equation}
[a_b^i,a_c^{k\dagger}]=\delta^{ik}_{bc}, \quad
\{f_b^{\rho},f_c^{\sigma\dagger}\}=\delta_{bc}^{\rho\sigma}, \quad
\rho, \sigma = 1,2 ,
\label{aaff}
\end{equation}
obeying the canonical (anti)commutation relations
\begin{equation}
[x_b^i,p_c^k] = i\delta^{ik} \delta_{bc} , \quad
\{\psi_b^{\alpha},\psi_c^{\beta}\} = \delta^{\alpha\beta} \delta_{bc}.
\label{acom}
\end{equation}
As usual bosonic variables are given  by
\begin{equation}
x^i_b = {1\over\sqrt{2}}(a^i_b+a_b^{i \dagger}), \quad
p^i_b = {1\over i \sqrt{2}}(a^i_b-a_b^{i \dagger}).  \label{XPD4}
\end{equation}
For fermionic variables we use the following representation for
a quantum Hermitian Majorana spinor
\begin{equation}
\psi_a={1+i\over 2\sqrt{2}}
   \left(\begin{array}{c}
   -   f_a^{1} - i f_a^{2} + i f_a^{1\dagger} +   f_a^{2\dagger} \\
   + i f_a^{1} -   f_a^{2} -   f_a^{1\dagger} + i f_a^{2\dagger} \\
   -   f_a^{1} + i f_a^{2} + i f_a^{1\dagger} -   f_a^{2\dagger} \\
   -i  f_a^{1} -   f_a^{2} +   f_a^{1\dagger} + i f_a^{2\dagger} \\
   \end{array} \right),   \label{eqmajor}
\end{equation}
which is easily shown to satisfy Eq.\ (\ref{acom}), with the help of
Eq.\ (\ref{aaff}). Other choices of fermionic creation and
annihilation operators are possible \cite{SF,HS,SMK}.

\subsection{The cutoff}

For completeness, we shortly review the practical construction of
the cut Fock space used in Ref.\ \cite{JWNP}. The entire Hilbert
space is generated by all independent polynomials of the {\em
elementary\/} creation operators $a_b^{i\dagger}$ and
$f_c^{\sigma\dagger}$ acting on the empty state, i.e., the state
with zero occupation number for all of the above-defined
oscillators. In practical applications we shall work in the
finite-dimensional Hilbert space of states containing a total of
at most $\nBmax$ bosonic quanta, i.e.,
\begin{equation}
n_B \le \nBmax, \quad
n_B \equiv \sum_{i,b} a_b^{i\dagger} a_b^i .
\label{eq:cutoff}
\end{equation}
There is no need to cut the fermionic number, which is limited to 6
by construction.

The physical Hilbert space is restricted to gauge-invariant states
only. It can be conveniently generated by all independent
polynomials of {\em gauge-invariant creators } -- bilinear or
trilinear combinations of $a^\dagger$'s and $f^\dagger$'s (the
explicit form will be given later).

Finally, since elementary creation and annihilation operators have a
straightforward action in the occupation-number basis, one can readily
calculate all matrix elements of the Hamiltonian and other
observables.

All these steps can be implemented automatically in a computer algebra
system. The matrix elements of any operator are calculated by writing
the operator in terms of creation and annihilation operators.
Finally, the complete spectrum and eigenstates of the cut Hamiltonian
(\ref{eq:Hamiltonian}) are obtained by numerical diagonalization.

This approach has proved reasonably successful. But of curse there is
a limit to it.  It is possible to improve the results considerably by
exploiting fully the symmetries of the cut system, and by foregoing
an explicit construction of states using only matrix elements
(cf.\ Ref.\ \cite{CW}).

\subsection{The symmetries}

Some of the symmetries were already exploited earlier to reduce the
size of the bases.  We now discuss shortly their significance.

The fermion number $n_F$ is conserved:
\begin{equation}
[n_F,H]=0, \quad n_F=f_a^{i\dagger}f_a^{i} .
\end{equation}
This is best seen in the Weyl representation of Dirac matrices,
where the Majorana spinor \ref{eqmajor} assumes the simple form
\cite{WBG}
\begin{equation}
\psi_W^T=(f_2^{\dagger},-f_1^{\dagger},f_1,f_2). \label{eqweyl}
\end{equation}
Since the Dirac $\alpha$ matrices are block-diagonal in this
representation, the fermionic Hamiltonian $H_F$
contains only bilinears of the type $f^{\dagger} f$. Therefore it
cannot change $n_F$.  Because the Pauli principle allows only six
colored Majorana fermions in this system, the whole Hilbert space
splits into seven sectors, $n_F=0,1,\dots,6$. The cutoff on the
bosonic quanta preserves $n_F$, and consequently the diagonalization
described above can be carried out independently in each fermionic
sector for finite $\nBmax$.

The system is invariant under the particle-hole symmetry
\begin{equation}
n_F \leftrightarrow 6 - n_F,  \label{ph}
\end{equation}
therefore it suffices to find the spectrum only in the first four
sectors, $n_F=0,1,2,3$, with the $n_F=3$ sector being selfdual
with respect to Eq.\ (\ref{ph}).

The local gauge invariance of the full (non-reduced) theory
turns into a global constraint of the reduced quantum mechanics.
Namely, the physical Hilbert space consists of the gauge-invariant
states, which in this case are invariant under the global SU(2)
rotations in the color space. This is taken care of by using the
gauge invariant combinations of the creation operators.  This symmetry
is preserved by the gauge invariant cutoff (\ref{eq:cutoff}), and was
already maximally exploited by working exclusively in the color-singlet
sector.

On the other hand, rotational invariance had not been fully used
until now.  Again, the cutoff is rotationally invariant and,
accordingly, only exactly degenerate SO(3) multiplets with
well-defined angular momentum were observed in the spectrum.
However no attempt was made to generate separate bases in each
angular momentum channel. This is the main source of improvement
in the present work and will be discussed in detail in the next
Section.

The system is also invariant under parity. In the $F=0$ sector, it
is equivalent to bosonic parity, $(-1)^{n_B}$, and states can be
classified according to the parity of $n_B$.

Finally, supersymmetry is broken by limiting $n_B$, since the
generators (\ref{QD4}) do change the number of bosonic quanta.  It is
therefore interesting to look for the restoration of supersymmetry
with the increase of the cutoff.  Indeed this was qualitatively
observed earlier.  Present improvements reveal the supersymmetric
spectrum with much better precision.

\subsection{Early results}

The $f^{\dagger} f$ structure of the fermionic Hamiltonian has an
instructive consequence. The interaction term $H_F$ vanishes in purely
bosonic sector $n_F=0$, which means that the effective Hamiltonian in
this sector is just the pure Yang-Mills, zero-volume Hamiltonian which
provides the starting point of the small volume expansion \cite{L}.  Indeed the
lowest eigenenergies found in this sector agree very well with
well established results of Ref.\ \cite{LM}. Later on, this test was extended
to higher states crosschecking with recent results by van Baal to
15-digit precision \cite{vBN,VBhp}.

Sizes of bases which were reached
in Refs.\ \cite{JWNP,KW} are quoted in Table \ref{bastab}.  They contain all angular momenta up
to $j_{\max} \le n_B + \half n_F$, also shown in the Table.  Due to the
particle-hole symmetry the structure in the $n_F=4,5,6$ sectors is
identical with that in $n_F=2,1,0$ respectively.  It will be
interesting to compare Table \ref{bastab} with our new results displayed in
Table \ref{tab:size-vs-nB}.

\begin{table}[tbp]
\begin{center}
\begin{tabular}{cccccc} \hline\hline
  $n_F$        &  $ 0 $ & $ 1 $ & $ 2 $ & $ 3 $  & \\  \hline
  $n_B$ &  $ N_{s}$\ \ \ $  \Sigma$ \ \  & $ N_{s} \ \ \  \Sigma$ \ \  &  $ N_{s} \ \ \  \Sigma$ \ \  & $ N_{s}\ \ \   \Sigma$  \ \ &  $\Sigma_B - \Sigma_F$ \\
   \hline
  0 &    1 \     1 \ \   &  -  \    -   \  &   1 \     1  \  &    4 \     4  \ & 0  \\
  1 &   -  \     1 \ \   &   6 \     6  \  &   9 \    10  \  &    6 \    10  \ & 0  \\
  2 &    6 \     7 \ \   &   6 \    12  \  &  21 \    31  \  &   42 \    52  \ & 0  \\
  3 &    1 \     8 \ \   &  36 \    48  \  &  63 \    94  \  &   56 \   108  \ & 0  \\
  4 &   21 \    29 \ \   &  36 \    84  \  & 111 \   205  \  &  192 \   300  \ & 0  \\
  5 &    6 \    35 \ \   & 126 \   210  \  & 240 \   445  \  &  240 \   540  \ & 0  \\
  6 &   56 \    91 \ \   & 126 \   336  \  & 370 \   815  \  &  600 \  1140  \ & 0  \\
  7 &   21 \   112 \ \   & 336 \   672  \  & 675 \  1490  \  &  720 \  1860  \ & 0  \\
  8 &  126 \   238 \ \   & 336 \  1008  \  & 960 \  2450  \  & 1500 \  3360  \ & 0  \\
\hline
 $j_{\max}$ &  8          &    17/2         &     9           &   19/2         &    \\
   \hline\hline
\end{tabular}
\end{center}
\caption{Sizes of bases generated in each fermionic sector,
  $n_F$. $N_s$ is the number of basis vectors with given number of
  bosonic quanta, $n_B$, while $\Sigma$ gives the cumulative size up to
  $n_B$. The last column gives the difference between the total number
  of the bosonic and fermionic states in all seven sectors.}
\label{bastab}
\end{table}

\begin{figure}[tbp]
\begin{center}
\epsfig{width=14cm,height=8cm,file=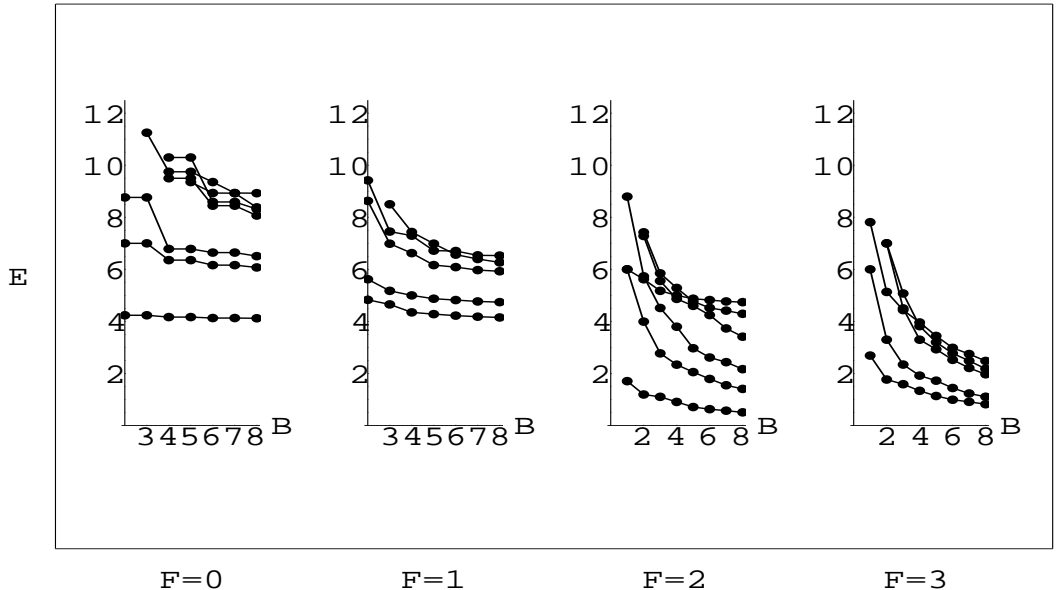}
\end{center}
\caption{Spectrum of $D=4$ SYMQM obtained in Ref.\ \cite{KW}.}
\label{fig:oldspectrum}
\end{figure}

In Fig.\ \ref{fig:oldspectrum} we display the lowest eigenenergies as
a function of the cutoff in all fermionic sectors.
Clearly the cutoff dependence is different in the
$n_F=0,1$ than in $n_F=2$ and $3$ sectors. Based on the experience
with simpler models, where the correlation between the nature of the
spectrum and the rate of convergence with $N_{cut}$ was established,
it was claimed that the spectrum in the $n_F=0,1,5,6$ sectors is
discrete, while it is continuous in the ``fermion rich'' sectors with
$n_F=2,3$ and $4$.  Recent analytic solutions of a sample of quantum
mechanical problems in a cut Hilbert space have proven that indeed
continuous spectra are characterized by the slow, power-like
dependence on the cutoff \cite{TW}. All these early results provided
an evidence that sizes of the bases displayed in Table \ref{bastab}
were sufficient to calculate
lowest localized states with a reasonable precision.

By computing directly supersymmetric images of lowest eigenstates it
was found that SUSY in the cutoff system was broken on the level of 10
-- 20 \% .  This was also confirmed by the Witten index
calculation.

\begin{figure}[tbp]
\begin{center}
\epsfig{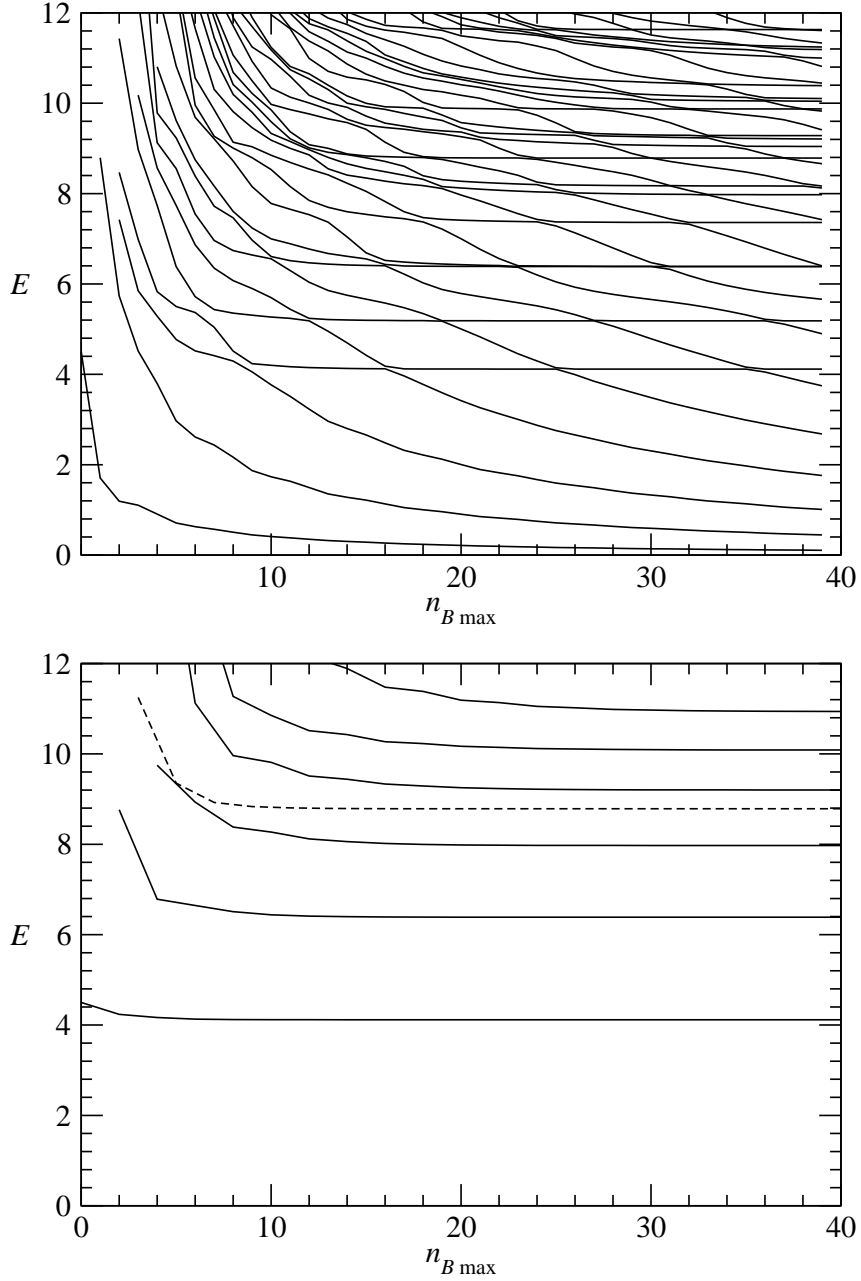}
\end{center}
\caption{High cutoff results from van Baal approach for two
available channels: $n_F=2,j=0$ (upper) and $n_F=0,j=0$ (lower).
The dashed line for $n_F=0$ is the only odd-parity level in this
energy range.} \label{fig:vanBaal}
\end{figure}

\subsection{Separation of variables}

The above conclusions, about the signature and coexistence of the
discrete and continuous spectra, have been dramatically confirmed
recently by van Baal \cite{VBhp}. Decomposing the solutions of the
nine-dimensional Schrodinger equation, in the $n_F=0, j=0$ and $n_F=2,
j=0$ channels, into covariant tensors, the problem was reduced to a
numerically affordable set of coupled ordinary differential equations.
 As a consequence, van Baal was able to push a cutoff
up to $\nBmax=39$ in these two channels, as shown in Fig.\
\ref{fig:vanBaal}.  The discrete, localized states with $n_F=0$
can be clearly seen with a very high precision.  Moreover, the
intricate nature of the $n_F=2$ sector is also evident.  As
expected, the localized states have quickly convergent
eigenenergies while the continuous spectrum manifests itself as a
family of levels which slowly fall with the cutoff. We postpone
the detailed discussion of this beautiful result until the global
picture of the solutions in all channels becomes clear.

Let us move now to the main subject of this paper which extends the
results just presented.  The new method allows to reach cutoffs in the
range $ 18 < \nBmax < 23 $ in all fermionic sectors and for all
angular momenta, providing at the same time detailed information on
the supersymmetric interrelations between eigenstates.

\section{The new approach}
\label{sec:newalgo}

We first present the basic features of the new algorithm, which allows
us to push the computation much further.

Rotational symmetry is exploited fully: all the objects in the
computation, beside being gauge singlets, belong to irreducible
representation of the rotation group Spin(3); this allows heavy use of
the traditional machinery of Clebsch-Gordan coefficients and $3j$ and
$6j$ symbols. (In the following, several formulae will be used; they
are reported in Appendix \ref{app:6j}).  In addition, parity symmetry
is used whenever possible.

Vectors are never constructed explicitly; we build instead a recursive
chain of identities between matrix elements of operators; this follows
closely our algorithm for the $D=2$ case \cite{CW}.

\subsection{Gauge-invariant operators with definite angular momentum
for the bosonic sector}
\label{sec:bosonic}

To avoid possible confusion, let us rename the bosonic creation and
annihilation operators defined in Eq.\ (\ref{aaff})
${\hat a}^{i\dagger}_b$ and ${\hat a}^{i}_b$ respectively.  In order
to create states of a definite total angular momentum
$\bm{J}^2 = j(j+1)$ and $J_z = m$, take the combinations
\begin{equation}
a^{\pm1\dagger}_b = \mp{1\over\sqrt2}
   \bigr({\hat a}^{1\dagger}_b\pm i{\hat a}^{2\dagger}_b\bigl),\quad
a^{0\dagger}_b = {\hat a}^{3\dagger}_b,\quad
a^{\pm1}_b = \mp{1\over\sqrt2}
   \bigr({\hat a}^1_b\mp i{\hat a}^2_b\bigl),\quad
a^0_b = {\hat a}^3_b,
\label{eq:am,ahi}
\end{equation}
so that $a^{m\dagger}_b|0\rangle$ is a state of angular momentum
$(1,m)$; now define ${\tilde a}^m_b = (-1)^{1+m} a^{-m}_b$: the new
creation and annihilation operators satisfy the canonical commutation
rules
\begin{equation}
[a^{m_1}_b, a^{m_2\dagger}_c] = \delta^{m_1m_2}\delta_{bc},\qquad
[{\tilde a}^{m_1}_b, a^{m_2\dagger}_c] =
\sqrt{3}\,C^{1\,1\,0}_{m_1m_20}\delta_{bc},
\end{equation}
$a^{m\dagger}_b$ and ${\tilde a}^m_b$ transform as spin-1 triplets
under rotations; they have odd parity $(-1)^{n_B}$.  (Here and in the
following, ${}^\dagger$ denotes the usual Hermitian conjugation
applied to a single $J_z$ component of an operator; e.g.,
$a^{1\dagger}_b$ is the Hermitian conjugate of $a^1_b$.)

{}From $a^m_b$ it is possible to build the bilinear gauge-invariant
operators $a^{m_1}_b a^{m_2}_b$, which are then decomposed in
components of given angular momentum $A_{j,m}$; let us introduce the
notation
\begin{equation}
\bigl(R_{j_1},S_{j_2} \bigr)_{j,m} \equiv \sum_{m_1,m_2}
   C^{j_1\,j_2\,j}_{m_1m_2m} R_{j_1,m_1} S_{j_2,m_2},
\label{eq:bracket}
\end{equation}
where $R$ and $S$ are arbitrary operators with definite rotational
properties; Eq.\ (\ref{eq:bracket}) implies
\begin{equation}
R_{j_1,m_1} S_{j_2,m_2} = \sum_{j,m}
    C^{j_1\,j_2\,j}_{m_1m_2m}  \bigl(R_{j_1},S_{j_2} \bigr)_{j,m}.
\label{eq:CGunfold}
\end{equation}

We can now define
\begin{equation}
A^\dagger_{j,m} = \bigl(a^\dagger_b, a^\dagger_b \bigr)_{j,m}; \qquad
{\tilde A}_{j,m} = (-1)^{j+m} \, A_{j,-m} =
\bigl({\tilde a}_b, {\tilde a}_b \bigr)_{j,m},
\end{equation}
where $A^\dagger_{j,m}$
is the Hermitian conjugate of $A_{j,m}$.  Since $A$ is a symmetric
combination of $a$'s, it has no $j{=}1$ components, but only 1
$j{=}0$ component and 5 $j{=}2$ components; $A^\dagger_{2,m}$ and
${\tilde A}_{2,m}$ transform as spin-2 quintets under rotations.

In order to express the commutation rules between $A$ and $A^\dagger$,
it is necessary to introduce the gauge-invariant ``mixed'' operators
\begin{equation}
B_{j,m} = -\half \bigl[\bigl({\tilde a}_b,a^\dagger_b\bigr)_{j,m}
     + (a^\dagger_b, {\tilde a}_b)_{j,m} \bigr]
= (-1)^{j+m} B^\dagger_{j,-m};
\end{equation}
in addition to 1 $j{=}0$ component and 5 $j{=}2$ components, $B$ has
also 3 $j{=}1$ components. We can now write
\begin{eqnarray}
[{\tilde A}_{j_1,m_1},A^\dagger_{j_2,m_2}] &=&
\sum_{j,m} c_{j_1j_2j} \, C^{j_1\,j_2\,j}_{m_1m_2m} \, B_{j,m};
\label{eq:AAdc} \\
{}[B_{j_1,m_2},A^\dagger_{j_2,m_2}] &=&
\sum_{j,m} c'_{j_1j_2j} \, C^{j_1\,j_2\,j}_{m_1m_2m} \, A^\dagger_{j,m}.
\label{eq:BAdc} \\
{}[B_{j_1,m_2},B_{j_2,m_2}] &=&
\sum_{j,m} c''_{j_1j_2j} \, C^{j_1\,j_2\,j}_{m_1m_2m} \, B_{j,m}.
\label{eq:BBc}
\end{eqnarray}
It would be pointless to write the detailed form of the coefficients
$c$, $c'$, and $c''$; their computation will be discussed in Appendix
\ref{app:algo}.

We also introduce the trilinear gauge-invariant creation
and annihilation operators
\begin{equation}
{\bar A}^\dagger = \varepsilon_{bcd}
{\hat a}^{\dagger\,1}_b {\hat a}^{\dagger\,2}_c {\hat a}^{\dagger\,3}_d =
i \varepsilon_{bcd}
(\bigl(a^\dagger_b, a^\dagger_c\bigr)_1,a^\dagger_d\bigr)_{0,0}, \qquad
\Abartilde = {\bar A}  \label{eq:AAAdag}
\end{equation}
(the notation $((R_{j_1},S_{j_2})_{j_3},T_{j_4})_{j,m}$ follows from
applying Eq.\ (\ref{eq:bracket}) twice), which have only the scalar
(i.e., spin-0) component, and the ``mixed'' trilinear operators
\begin{equation}
{\bar B}^\dagger_{j,m} = \varepsilon_{bcd}
(\bigl(a^\dagger_b, a^\dagger_c\bigr)_1,{\tilde a}_d\bigr)_{j,m},
\quad \Bbartilde_{j,m} = - \varepsilon_{bcd}
(\bigl({\tilde a}_b, {\tilde a}_c\bigr)_1,a^\dagger_d\bigr)_{j,m},
\quad {\bar B}_{j,m} = (-1)^{j+m} \Bbartilde_{j,m}.
\end{equation}

The above-defined operators form a complete set of gauge-invariant
bosonic operators, in the sense that any gauge-invariant bosonic
operator can be written as a polynomial in these operators.  In
particular, we can write $H_K$ and $H_P$ in terms of $A^\dagger$,
$\tilde A$, and $B$ as
\begin{eqnarray}
H_K &=& \fourth\sqrt{3}\bigl(A^\dagger_{0,0} +
    {\tilde A}_{0,0} - 2 B_{0,0}\bigr);
\label{eq:HK} \\
{4\over g^2} H_P &=& \fourth(A^\dagger,A^\dagger)_P
+ \fourth({\tilde A},{\tilde A})_P
+ (A^\dagger,B)_P + (B,{\tilde A})_P
+ (B,B)_P + \half(A^\dagger,{\tilde A})_P \nonumber \\
&&+\, \sqrt{3}\bigl(A^\dagger_{0,0} + {\tilde A}_{0,0} + B_{0,0}\bigr),
\label{eq:HP} \\
(R,S)_P &\equiv& 2 R_{0,0}S_{0,0} - \sqrt{5} (R_2,S_2)_{0,0}. \nonumber
\end{eqnarray}

\subsection{Fermionic operators with definite angular momentum}
\label{sec:fermionic}

To identify fermionic creation operators with definite angular
momentum, recall the origin of the parametrization
(\ref{eqmajor}). It represents a Majorana fermion in Majorana
representation of Dirac matrices and was obtained by a unitary
transformation of a Majorana fermion in the Weyl representation
(\ref{eqweyl}) \cite{JWNP}.  Therefore $f^{\sigma\dagger}_b$
creates in fact a fermion in the Weyl representation and as such
carries definite angular momentum. This follows from the explicit
form of the spin operator $S^3$ defined in Eq.\ (\ref{JD4}):
\begin{equation}
S^3 = {1\over 2}\psi^T_a\Sigma^{12}\psi_a =
{1\over 2} \left(f^{1\dagger}_bf^{1}_b - f^{2\dagger}_b f^{2}_b \right),
\end{equation}
which can be obtained in either Weyl or Majorana representations of
Dirac matrices.  Therefore, ${\hat f}_b^\sigma$, the fermionic
creation and annihilation operators defined in Eq.\ (\ref{aaff}), are
already the desired operators, and we set
\begin{eqnarray*}
f^{\fh1}_b = {\hat f}^1_b, &\qquad& f^{-\fh1}_b = {\hat f}^2_b; \\
{\tilde f}^m_b = (-1)^{\fh1+m} f^{-m}_b &\qquad&
({\tilde f}^{\fh1}_b = -f^{-\fh1}_b, \quad {\tilde f}^{-\fh1}_b = f^{\fh1}_b);
\end{eqnarray*}
$f^{m\dagger}_b$ and ${\tilde f}^m_b$ are $j=\half$, $J_z=m$
operators; the anticommutation relations are
\[ \{f^{m_1}_b,f^{m_2\dagger}_c\} = \delta_{m_1m_2}\delta_{bc}, \qquad
   \{{\tilde f}^{m_1}_b,f^{m_2\dagger}_c\} =
      -\sqrt{2}\,C^{\fh1\,\fh1\,0}_{m_1m_20}\,\delta_{bc}. \]

\subsection{Gauge-invariant operators involving fermions}
\label{sec:GI-ferm}

Let us complete the set of gauge-invariant operators, bilinear
${\cal O}^{n_F,n_B}_{j,m}$ or trilinear
${\bar{\cal O}}^{n_F,n_B}_{j,m}$, with definite $\bm{J}^2$, $J_z$,
$n_F$, and $n_B$.  In the bilinear case, we complement the bosonic
operators ${\cal O}^{0,2}_{j,m}=A^\dagger_{lm}$, ${\tilde A}$, and
${\cal O}^{0,0}_{j,m}=B_{lm}$ with
\begin{eqnarray*}
{\cal O}^{1,1}_{j,m} = F^\dagger_{j,m} =
     \bigl(f^\dagger_b,a^\dagger_b\bigr)_{j,m}, &\quad&
{\tilde F}_{j,m} = (-1)^{j+m} F_{j,-m}, \quad
j=\txf12,\txf32; \\
{\cal O}^{1,-1}_{j,m} = \Theta^\dagger_{j,m} =
     -\bigl(f^\dagger_b,{\tilde a}_b\bigr)_{j,m}, &\quad&
{\tilde \Theta}_{j,m} = (-1)^{j+m} \Theta_{j,-m}, \quad
j=\txf12,\txf32; \\
{\cal O}^{\prime\,0,0}_{j,m} = \Phi_{j,m} =
     -\bigl(f^\dagger_b,{\tilde f}_b\bigr)_{j,m}, &\quad&
\Phi_{j,m} = (-1)^m \Phi^\dagger_{j,-m}, \quad j=0,1 \\
{\cal O}^{2,0}_{0,0} = G^\dagger_{0,0} =
     \bigl(f^\dagger_b,f^\dagger_b\bigr)_{0,0}, &\quad&
{\tilde G}_{0,0} = G_{0,0}, \quad j=0
\end{eqnarray*}
($G^\dagger_{1,m}$ vanishes identically); note that $F$, $G$,
$\Theta$, and $\Phi$ give zero when applied to a bosonic state.

In the trilinear case, we complement the bosonic creation and
annihilation operators ${\bar{\cal O}}^{0,3}_{0,0} = {\bar A}^\dagger$
and ${\bar A}$ with
\begin{eqnarray*}
{\bar{\cal O}}^{1,2}_{j,m} = {\bar F}^\dagger_{j,m} &=& i\varepsilon_{bcd}
   \bigl(f^\dagger_b,(a^\dagger_c,a^\dagger_d)_1\bigr)_{j,m}, \qquad
   \Fbartilde_{j,m} = (-1)^{j+m} {\bar F}_{j,-m}, \quad
j=\txf12,\txf32; \\
{\bar{\cal O}}^{2,1}_{j,m} = {\bar G}^\dagger_{j,m} &=& i\varepsilon_{bcd}
   \bigl((f^\dagger_c,f^\dagger_d)_1,a^\dagger_b\bigr)_{j,m}, \qquad
\Gbartilde_{j,m} = (-1)^{j+m} {\bar G}_{j,-m}, \quad
j=0,1,2; \\
{\bar{\cal O}}^{3,0}_{\txf32,m} =
   {\bar I}^\dagger_{\txf32,m} &=& i\varepsilon_{bcd}
   \bigl((f^\dagger_c,f^\dagger_d)_1,f^\dagger_b\bigr)_{\txf32,m}, \qquad
   \Ibartilde_{j,m} = (-1)^{j+m} {\bar I}_{j,-m}, \quad
j=\txf32;
\end{eqnarray*}
the antisymmetrized product of two $a^\dagger$'s only produces $j=1$,
and likewise for $f^\dagger$'s; the factor $i$ in ${\bar F}^\dagger$,
${\bar G}^\dagger$, and ${\bar I}^\dagger$ is required to have real
matrix elements of $H$ between a $n_B$ even, $n_F=1$ and a $n_B$ odd,
$n_F=1$ state. We also define the ``mixed'' operators
${\bar{\cal O}}^{0,1} = {\bar B}^\dagger$, $\Bbartilde$, and
\begin{eqnarray*}
{\bar{\cal O}}^{1,0}_{j,m} = {\bar\Omega}^\dagger_{j,m} &=& -i\varepsilon_{bcd}
   \bigl((f^\dagger_b,f^\dagger_c)_1,{\tilde f}_b\bigr)_{j,m}, \qquad
\Omegabartilde_{j,m} = (-1)^{j+m} {\bar\Omega}_{j,-m}, \quad
{\textstyle j=\fh1,\fh3}; \\
{\bar{\cal O}}^{0,1;j'}_{j,m} = {\bar\Phi}^{\dagger\,j'}_{j,m} &=&
-i\varepsilon_{bcd}
   \bigl((f^\dagger_b,{\tilde f}_c)_{j'},a^\dagger_d\bigr)_{j,m}, \qquad
   \Phibartilde^{j'}_{j,m} = (-1)^{j+m} {\bar\Phi}^{j'}_{j,-m},\quad
j'=0,1,\;j=0,1,2; \\
{\bar{\cal O}}^{2,-1}_{j,m} = {\bar\Xi}^\dagger_{j,m} &=& -i\varepsilon_{bcd}
   \bigl((f^\dagger_b,f^\dagger_c)_1,{\tilde a}_d\bigr)_{j,m}, \qquad
\Xibartilde_{j,m} = (-1)^{j+m} {\bar\Xi}_{j,-m}, \quad
j=0,1,2; \\
{\bar{\cal O}}^{1,0;j'}_{j,m} = {\bar\Psi}^{\dagger\,j'}_{j,m} &=&
   -i\varepsilon_{bcd}
   \bigl(({\tilde a}_b,a^\dagger_c)_{j'},f^\dagger_d\bigr)_{j,m}, \qquad
   \Psibartilde^{j'}_{j,m} = (-1)^{j+m} {\bar\Psi}^{j'}_{j,-m}, \quad
{\textstyle j'=0,1,2,\;j=\fh1,\fh3,\fh5;} \\
{\bar{\cal O}}^{1,-2}_{j,m} = {\bar\Theta}^\dagger_{j,m} &=&
   i\varepsilon_{bcd}
   \bigl(({\tilde a}_b,{\tilde a}_c)_1,f^\dagger_d\bigr)_{j,m}, \qquad
\Thetabartilde_{j,m} = (-1)^{j+m} {\bar\Theta}_{j,-m}, \quad
{\textstyle j=\fh1,\fh3};
\end{eqnarray*}
note that ${\bar\Omega}$, ${\bar\Omega}^\dagger$, ${\bar\Phi}$,
${\bar\Phi}^\dagger$, ${\bar\Xi}$, ${\bar\Psi}$, and ${\bar\Theta}$
give zero when applied to a bosonic state.

We can establish (anti)commutation relations between pairs of
gauge-invariant operators, similar to Eqs.\ (\ref{eq:AAdc}) and
(\ref{eq:BAdc}); it would be pointless to present here their explicit
form; their computation will be discussed in Appendix \ref{app:algo}.

The above-defined operators form a complete set of gauge-invariant
operators, in the sense that any gauge-invariant operator can be
written as a polynomial in these operators.  In particular, we show by
explicit computation that
\begin{equation}
H_F = -g\sqrt{3} \bigl({\bar\Phi}^{\dagger\,1}_{0,0}
+ {\bar\Phi}^1_{0,0}\bigr).
\label{eq:HF}
\end{equation}

$Q$ must be decomposed in components with definite $J_z$ and $n_F$;
since in our Majorana representation $\Gamma^j$ is real and
$\Sigma^{jk}$ is purely imaginary, $Q_\alpha$ is Hermitian.  Denoting
the $n_F=1$ and $n_F=-1$ doublets by $Q^\dagger_m$ and ${\tilde Q}_m$
respectively, we have
\begin{eqnarray}
Q^\dagger_{\fh1} &=& \txf12 e^{i\theta} (Q_1 - i Q_2 + Q_3 + i Q_4),
\nonumber \\
Q^\dagger_{-\fh1} &=& \txf12 e^{i\theta} (i Q_1 + Q_2 - i Q_3 + Q_4),
\nonumber \\
{\tilde Q}_m &=& (-1)^{\fh1+m} Q_{-m}, \label{WeylQ}
\end{eqnarray}
where $\theta$ is an arbitrary phase; the anticommutation relations are
\begin{eqnarray*}
&\{Q^\dagger_m,Q^\dagger_{m'}\} = 0, \qquad
 \{{\tilde Q}_m,{\tilde Q}_{m'}\} = 0, \\
&\bigl({\tilde Q},Q^\dagger\bigr)_{0,0} +
 \bigl(Q^\dagger,{\tilde Q},\bigr)_{0,0} = \fourth\sqrt{2}\,H, \\
&\bigl({\tilde Q},Q^\dagger\bigr)_{1,m} +
 \bigl(Q^\dagger,{\tilde Q},\bigr)_{1,m} \equiv v_m =
 \sqrt{2}\,x^m_b G_b,
\end{eqnarray*}
where $x^m_b$ is defined by the analogous of Eq.\ (\ref{eq:am,ahi});
$v_m$ gives zero when applied to a gauge-invariant state; the only
nontrivial anticommutator can be rewritten as
\begin{equation}
H = \txf14 \sum_m \{Q_m,Q^\dagger_m\};
\label{eq:QH}
\end{equation}
we choose $\theta=-\txf14\pi$; an explicit computation gives
\begin{equation}
Q^\dagger_m =
  \sqrt{\txf32}\bigl(F^\dagger_{\fh1m} - \Theta^\dagger_{\fh1m}
  - \txf12g{\bar F}^\dagger_{\fh1m} + \txf12g{\bar \Theta}^\dagger_{\fh1m}
  + g{\bar\Psi}^{\dagger\,1}_{\fh1m}\bigr).
\label{eq:Q-def}
\end{equation}
Note that, with the present conventions, all matrix elements of
interest are real.

\subsection{Construction and orthonormalization of states
with definite angular momentum}
\label{sec:ortho}

All states are classified into {\em even\/} and {\em odd\/} states,
according to the parity of $p \equiv n_F+n_B\ (\mod 2)$.  (This label
coincides with parity only for $n_F=0$ states.)

It is useful to set up a common naming scheme for all our creation
operators: $X(\nu,p)^\dagger$ is the creation operator with $n_F=\nu$
and $n_B=2+p-\nu$; i.e., $X(0,0)=A$, $X(1,0)=F$, $X(2,0)=G$,
$X(0,1)={\bar A}$, $X(1,1)={\bar F}$, $X(2,1)={\bar G}$, and
$X(3,1)={\bar I}$; $X(3,0)$ is identically zero.

We build our states recursively, applying $X(\nu,p)^\dagger$ to a
state of an orthonormal basis with definite $\bm{J}^2$ and $J_z$, and
taking linear combinations to produce again an orthonormal basis.

It is important to note that, given the $\varepsilon$ contraction rule
\begin{equation}
\varepsilon_{i_1i_2i_3}\,\varepsilon_{j_1j_2j_3} =
\det\Vert\delta_{i_\alpha j_\beta}\Vert,
\label{eq:epsilon}
\end{equation}
the product of two trilinear operators can always be decomposed into a
sum of products of three bilinear operators; therefore, even states
can be built by applying any number of even ($p=0$) creation operators
to the vacuum; correspondingly, odd states can be built by applying
one odd ($p=1$) and any number of even creation operators to the
vacuum.  ${\bar A}^\dagger$ is never needed in combination with
fermionic operators, since $X(\nu,0)^\dagger {\bar A}^\dagger$ for
$\nu>0$ can be written as a linear combination of terms of the form
$X(\nu,1)^\dagger A^\dagger$.

Creation operators (which (anti)commute between themselves) can be
ordered to get every fermionic operator to the left of every bosonic
operator and every trilinear operator to the left of every bilinear
operator.  Therefore, using the notation $|n_F,n_B\rangle$ for our
states, we can build all bosonic states from even bosonic states as
$|0,2n{+}p\rangle = X(0,p)^\dagger|0,2n{-}2\rangle$, and all fermionic
states of parity $p$ from even states of lower $n_F$ as
$|n_F,n_B\rangle =
X(\nu,p)^\dagger|n_F{-}\nu,n_B{-}2{-}p{+}\nu\rangle$, $1\le\nu\le
n_F$.

In order to create a fermionic state with $n_B>n_F+1$, at least one
$A^\dagger$ must be used; therefore, such states can also be built as
$|n_F,n_B\rangle = A^\dagger|n_F,n_B{-}2\rangle$; this second recipe
turns out to be much more efficient, both in generating and
orthonormalizing the states and in computing matrix elements of
operators between them.

A basis for the sector with given $n_F$ and $n_B$ is contained in the
set
\begin{equation}
|j,m,n_F,n_B;\nu,p,j_1,j_2,i\rangle = \sum_{m_1,m_2} C^{j_1\,j_2\,j}_{m_1m_2m}
X(\nu,p)^\dagger_{j_1,m_1} |j_2,m_2,n_F{-}\nu,n_B{-}2{-}p{+}\nu;i\rangle,
\label{eq:states-build}
\end{equation}
where
\begin{eqnarray}
\quad \nu=0,\; p \equiv n_F+n_B\ (\mod 2),
&\qquad& n_F=0; \nonumber \\
\quad 1\le\nu\le n_F,\; p \equiv n_F+n_B\ (\mod 2),
&\qquad& n_F>0,\; n_B \le n_F+1;
\label{eq:nu-choice} \\
\quad \nu = 0,\; p = 0,
&\qquad& n_F>0,\; n_B > n_F+1. \nonumber
\end{eqnarray}

The scalar product of two such states can be written as
\begin{eqnarray*}
&&\langle j',m',n'_F,n'_B;\nu',p',j'_1,j'_2,i'
  |j,m,n_F,n_B;\nu,p,j_1,j_2,i\rangle = \\
&&\qquad\qquad \delta_{j'j}\delta_{m'm}\delta_{n'_Fn_F}\delta_{n'_Bn_B}\,
S^{j,n_F,n_B}_{\nu',p',j'_1,j'_2,i';\nu,p,j_1,j_2,i}.
\end{eqnarray*}
By Gram-Schmidt orthonormalization we obtain the orthonormal basis
\begin{eqnarray}
|j,m,n_F,n_B;i\rangle
&=& \sum_{\nu,p,j_1,j_2,j} R^{j,n_F,n_B}_{i;\nu,p,j_1,j_2,j}
    |j,m,n_F,n_B;\nu,p,j_1,j_2,j\rangle,
\label{eq:orthoR} \\
&&\kern-3cm
   \sum_{\nu',p',j'_1,j'_2,j';\nu,p,j_1,j_2,j}
   R^{j,n_F,n_B}_{i';\nu',p',j'_1,j'_2,j'}
   S^{j,n_F,n_B}_{\nu',p',j'_1,j'_2,j';\nu,p,j_1,j_2,j}
   R^{j,n_F,n_B}_{i;\nu,p,j_1,j_2,j}
   = \delta_{i'i}. \nonumber
\end{eqnarray}
The states of the set (\ref{eq:states-build}) may not be linearly
independent; this is however not a serious problem: Gram-Schmidt
orthonormalization will select an orthonormal basis and give a
non-square matrix $R$.
Eq.\ (\ref{eq:orthoR}) implies
\begin{eqnarray}
&&  {\langle j,n_F,n_B;i'\Vert X(\nu,p)^\dagger_{j_1}
    \Vert j_2,n_F{-}\nu,n_B{-}2{-}p{+}\nu;i\rangle} \nonumber \\
&=& \sqrt{2j+1}
    \sum_{\nu',p',j'_1,j'_2,j'} R^{j,n_F,n_B}_{i';\nu',p',j'_1,j'_2,j'}
S^{j,n_F,n_B}_{\nu',p',j'_1,j'_2,i';\nu,p,j_1,j_2,i},
\label{eq:X-me1}
\end{eqnarray}
where $\langle\alpha'\Vert{\cal O}\Vert\alpha\rangle$ denotes a
reduced matrix element, cf.\ Appendix \ref{app:W-E}.

To compute the scalar product of two such states, define
\[ \{ {\tilde X}(\nu',p')_{j'_1,m'_1}, X(\nu,p)^\dagger_{j_1,m_1} \}_\pm
   = \sum_{j_3,m_3} C^{j_1'\,j_1\,j_3}_{m_1'm_1m_3}
   K^{(\nu',p',j_1';\nu,p,j_1)}_{j_3,m_3}, \]
where the sign is $+$ (anticommutator) when both $\nu$ and $\nu'$ are
odd, $-$ (commutator) otherwise.

Using completeness and applying well-known identities similar to Eqs.\
(\ref{eq:CGcompl}) and (\ref{eq:CG3}), we obtain
\begin{eqnarray}
&&  \langle j',m',n_F,n_B;\nu',p',j'_1,j'_2,i'
    |j,m,n_F,n_B;\nu,p,j_1,j_2,i\rangle \nonumber \\
&=& \mp\delta_{j'j}\delta_{m'm}\sum_{j_3,i_3}
    (-1)^{j_1+j'_2+j_3} \, \sixj{j&j'_1&j'_2 \\ j_3&j_1&j_2} \nonumber \\
&&\qquad\times\,
    {\langle j'_2,{\bar n}'_F,{\bar n}'_B;i'\Vert  X(\nu,p)^\dagger_{j_1}
    \Vert j_3,n''_F,n''_B;i_3\rangle} \;
    {\langle j_3,n''_F,n''_B;i_3\Vert  {\tilde X(\nu',p')_{j_1}}
     \Vert j_2,{\bar n}_F,{\bar n}_B;i\rangle} \nonumber \\
&+& \delta_{j'j} \delta_{m'm} \sum_{j_3} (-1)^{j-j_1+j_2+2j_3}
    \sixj{j&j'_1&j'_2 \\ j_3&j_2&j_1}
     {\langle j'_2,{\bar n}'_F,{\bar n}'_B;i'\Vert
     K^{(\nu',p',j'_1;\nu,p,j_1)}_{j_3}
     \Vert j_2,{\bar n}_F,{\bar n}_B;i\rangle},
\label{eq:sp-reduction}
\end{eqnarray}
where ${\bar n}_F$, ${\bar n}_B$, ${\bar n}'_F$, ${\bar n}'_B$,
$n''_F$, and $n''_B$ are fixed by the selection rules.  The r.h.s.\
involves matrix elements of operators between states with lower $n_F$
or $n_B$.

In the case $p=p'=1$, the (anti)commutators would involve all
trilinear operators; to avoid this, applying Eq.\ (\ref{eq:epsilon})
we rewrite ${\tilde X}(\nu',1) \, X(\nu,1)^\dagger$ as a sum of
products of three bilinear operators, decomposed in components of
definite angular momentum; they are dealt with exactly like the
commutator term in the above equation, with the same factors and $6j$
symbols.  The explicit computation of the decomposition will be
discussed in Appendix \ref{app:algo}.

\subsection{Recursive computation of matrix elements of operators}
\label{sec:recursive}

Our task is to compute a matrix element of the form
\[ \langle j',m',n'_F,n'_B;i'|{\cal O}_{j'',m''}|j,m,n_F,n_B;i\rangle, \]
where ${\cal O}$ is an operator with a definite number of fermionic
and bosonic quanta $n''_F$ and $n''_B$; let us take $n'_F=n_F+n''_F$,
$n'_B=n_B+n''_B$ (otherwise, the matrix element is zero).

Apply Eqs.\ (\ref{eq:states-build}) and (\ref{eq:orthoR}) to the ket:
\begin{eqnarray*}
&& \langle j',m',n'_F,n'_B;i'|
   {\cal O}_{j'',m''}|j,m,n_F,n_B;i\rangle \\
&=& \sum_{m_1,m_2;\nu,p,j_1,j_2,j}
    C^{j_1\,j_2\,j}_{m_1m_2m} R^{j,n_F,n_B}_{i;\nu,p,j_1,j_2,j} \\
&&\quad\times\,
    \langle j',m',n'_F,n'_B;i'| {\cal O}_{j'',m''}
    X(\nu,p)^\dagger_{j_1,m_1} |j_2,m_2,n_F{-}\nu,n_B{-}2{-}p{+}\nu;j\rangle.
\end{eqnarray*}
Using the (anti)commutator
\[ \{{\cal O}_{j_1,m_1}, X(\nu,p)^\dagger_{j_2,m_2} \}_\pm
   = \sum_{j_3,m_3} C^{j_1\,j_2\,j_3}_{m_1m_2m_3} \,
   K^{({\cal O},j_1;\nu,p,j_2)}_{j_3,m_3} \]
and completeness, we obtain
\begin{eqnarray*}
&& {\langle j',n'_F,n'_B;i'\Vert
   {\cal O}_{j''}\Vert j,n_F,n_B;i\rangle} \\
&=& \mp \sum_{\nu,p,j_1,j_2,j;j_3,i_3}
    (-1)^{j+j''+j_1+j_3} \, \sqrt{2j+1}
    \sixj{j&j''&j' \\ j_3&j_1&j_2}
    R^{j,n_F,n_B}_{i;\nu,p,j_1,j_2,j} \\
&&\quad\times\,
    {\langle j',n'_F,n'_B;i'\Vert  X(\nu,p)^\dagger_{j_1}
    \Vert j_3,n'_F{-}\nu,n'_B{-}2{-}p{+}\nu;i_3\rangle} \\
&&\quad\times\,
    {\langle j_3,n'_F{-}\nu,n'_B{-}2{-}p{+}\nu;i_3\Vert
    {\cal O}_{j''}\Vert ,j_2,n_F{-}\nu,n_B{-}2{-}p{+}\nu;j\rangle} \\
&&\ +\,\sum_{\nu,p,j_1,j_2,j;j_3}
    (-1)^{j'+j''+j_1+j_2} \,\sqrt{(2j+1)(2j_3+1)}
    \sixj{j&j''&j' \\ j_3&j_2&j_1}
    R^{j,n_F,n_B}_{i;\nu,p,j_1,j_2,j} \\
&&\quad\times\,
    {\langle j',n'_F,n'_B;i'\Vert
    K^{({\cal O},j'';\nu,p,j_1)}_{j_3,m_3}
    \Vert j_2,n_F{-}\nu,n_B{-}2{-}p{+}\nu;j\rangle}.
\end{eqnarray*}
In the case of $p=1$ and trilinear ${\cal O}$, we can again resort to
the use of Eq.\ (\ref{eq:epsilon}) to rewrite ${\cal O} \,
X(\nu,1)^\dagger$ as a sum of products of three bilinear operators,
decomposed in components of definite angular momentum.
Every matrix element is computed in terms of matrix
elements for smaller $n_F$ and/or $n_B$; the recursion is closed when
a matrix element is obviously zero, or when Eq.\ (\ref{eq:X-me1}) can
be applied; the only nontrivial case is
\[ {\langle0,0,0;1\Vert B_j\Vert 0,0,0;1\rangle} =
   -\txf32\sqrt{3}\,\delta_{j0}. \]

Finally, to compute $H$ apply Eqs.\ (\ref{eq:HK}), (\ref{eq:HP}),
and (\ref{eq:HF}); to compute $Q$, apply Eq.\ (\ref{eq:Q-def}).

The implementation of the algorithm will be described in Appendix
\ref{app:algo}.

\section{Results}

\subsection{Hilbert space: sectors, channels and diamonds}

\label{sec:diamonds}

The approach described in Sect.\ \ref{sec:newalgo} allows to deal with
a considerably larger Hilbert space than the direct method, cf.\
Tables \ref{bastab} and \ref{tab:size-vs-nB}.
\begin{table}[tbp]
\begin{center}
\begin{tabular}{|r|rr|rr|rr|rr|rr|}
\hline
$\nBmax$ &
\multicolumn{2}{c|}{$n_F=0$} &
\multicolumn{2}{c|}{$n_F=1$} &
\multicolumn{2}{c|}{$n_F=2$} &
\multicolumn{2}{c|}{$n_F=3$} &
\multicolumn{2}{c|}{total} \\
\hline
0 & 1 & 1 & 0 & 0 & 1 & 1 & 1 & 4 & 5 & 8 \\
1 & 0 & 0 & 2 & 6 & 3 & 9 & 2 & 6 & 12 & 36 \\
2 & 2 & 6 & 2 & 6 & 7 & 21 & 10 & 42 & 32 & 108 \\
3 & 1 & 1 & 8 & 36 & 15 & 63 & 13 & 56 & 61 & 256 \\
4 & 5 & 21 & 8 & 36 & 25 & 111 & 36 & 192 & 112 & 528 \\
5 & 2 & 6 & 22 & 126 & 44 & 240 & 44 & 240 & 180 & 984 \\
6 & 10 & 56 & 22 & 126 & 64 & 370 & 92 & 600 & 284 & 1704 \\
7 & 5 & 21 & 48 & 336 & 101 & 675 & 108 & 720 & 416 & 2784 \\
8 & 18 & 126 & 48 & 336 & 136 & 960 & 195 & 1500 & 599 & 4344 \\
9 & 10 & 56 & 92 & 756 & 199 & 1575 & 222 & 1750 & 824 & 6524 \\
10 & 30 & 252 & 92 & 756 & 255 & 2121 & 364 & 3234 & 1118 & 9492 \\
11 & 18 & 126 & 160 & 1512 & 354 & 3234 & 407 & 3696 & 1471 & 13440 \\
12 & 48 & 462 & 160 & 1512 & 438 & 4186 & 622 & 6272 & 1914 & 18592 \\
13 & 30 & 252 & 260 & 2772 & 584 & 6048 & 686 & 7056 & 2434 & 25200 \\
14 & 72 & 792 & 260 & 2772 & 704 & 7596 & 996 & 11232 & 3068 & 33552 \\
15 & 48 & 462 & 400 & 4752 & 910 & 10530 & 1086 & 12480 & 3802 & 43968 \\
16 & 105 & 1287 & 400 & 4752 & 1075 & 12915 & 1515 & 18900 & 4675 & 56808 \\
17 & 72 & 792 & 590 & 7722 & 1355 & 17325 & 1638 & 20790 & 5672 & 72468 \\
18 & 148 & 2002 & 590 & 7722 & 1575 & 20845 &  &  &  &  \\
19 & 105 & 1287 & 840 & 12012 &  &  &  &  &  &  \\
20 & 203 & 3003 & 840 & 12012 &  &  &  &  &  &  \\
21 & 148 & 2002 &  &  &  &  &  &  &  &  \\
22 & 272 & 4368 &  &  &  &  &  &  &  &  \\
23 & 203 & 3003 &  &  &  &  &  &  &  &  \\
\hline
\end{tabular}
\end{center}
\caption{Number of SO(3) multiplets (left) and size of the basis
(right) of the physical Hilbert space, at given $n_B$ for each
$n_F$ sector. Right columns correspond to the
 left ones of Table \ref{bastab}. The last column gives respective
 sizes summed over all seven fermionic sectors.  }
\label{tab:size-vs-nB}
\end{table}
With the recursive algorithm implemented in Mathematica we were able
to compute all matrix elements of $H$ and $Q$ on a single PC in a time
ranging from 2 minutes for $n_F=0$ alone to 140 hours for the whole
computation. The whole Hilbert space was effectively split into seven
sectors of fixed fermion number, $0\le n_F \le 6$, which in turn
decouple into channels of fixed angular momentum $j$. In Appendix
\ref{app:size} we quote sizes of bases in all $(n_F,j)$ channels for
all available values of $n_B$.

It is useful to represent this decomposition on a $(n_F,j)$ plane
where circles corresponding to the individual channels form a regular
mesh \footnote{With two symmetric dislocations: there are no states with
$j=1$ and $n_F=0,6$.}.
\begin{figure}[tbp]
\begin{center}
\epsfig{width=8cm,file=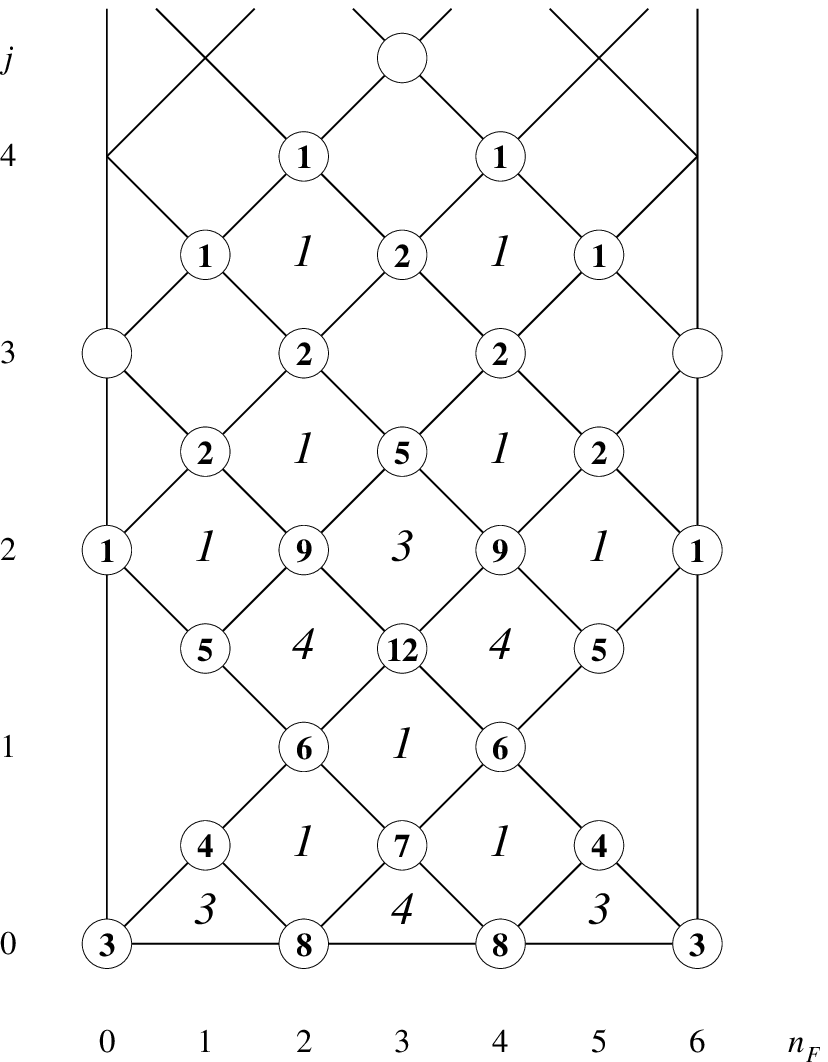}
\end{center}
\vskip-4mm
\caption{
 Populations of channels (bold) versus equivalent number of supermultiplets
 (italics) at $n_{B_{max}}=3$.}
\label{fig:diam.nB3}
\end{figure}
Fig.\ \ref{fig:diam.nB3} shows such a map together with the number SO(3) multiplets
 in each channel (for a particular value of a cutoff \nbm). The
distribution of states among channels is such that each SO(3)
multiplet belongs to one and only one diamond adjacent to the
vertex.  This "population" of all vertices determines the
 multiplicities of individual diamonds, i.e., the number of supermultiplets
  reproduced at given \nbm . The latter are
given in italic in the figure. More precisely, if $R_I$ denotes
the multiplicity of a diamond I, and $d_i$ is a number of SO(3) multiplets in
a channel $i=(n_F,j)$ then
\eq
d_i=\Sigma_{I|i} R_I,
\label{eq:d-vs-R}
\eqx
where $\Sigma_{I|i}$ means summation over I's adjacent to $i$.
These diamonds are nothing but supersymmetric multiplets which
will be discussed below in detail. At present stage they play only
a kinematical role --- they provide an alternative way of
classifying all basis states. Our cutoff \nbm\ violates
supersymmetry, however it violates it rather gently. Namely, for
every {\em odd} \nbm\ the total number of states is such that they fill
the integer number of supermultiplets.  Taking into account the SO(3)
degeneracy, we see immediately that the numbers of fermionic and
bosonic states in a diamond match. These last two properties
account for the exact balance between fermionic and bosonic states
found earlier, cf.\ the last column of Table \ref{bastab}.
For {\em even\/} \nbm, Eq.\ (\ref{eq:d-vs-R}) also holds, but exactly
two diamonds at highest $j$ have $d=-1$.
However, this does not spoil the exact balance, and it is irrelevant
in our perspective of increasing \nbm\ at fixed $j$.

We shall discuss now some detailed features
of the model which follow from supersymmetry and rotational
symmetry.

\subsection{The algebra of SUSY generators: supermultiplets}
\label{sec:SUSY-alg}

It is convenient to work with the Weyl generators defined in Eq.\
(\ref{WeylQ}), which carry definite fermionic number $n_F$ and
angular momentum ($m=\pm\half$), and satisfy the standard
anticommutation relations \eq \{Q_m,Q^{\dagger}_n\}=4 H
\delta_{mn},\quad \{Q^{\dagger}_m,Q^{\dagger}_n\}=\{Q_m,Q_n\}=0
\label{QQcom} \eqx in the gauge-invariant sector.

\begin{figure}[tb]
\begin{center}
\epsfig{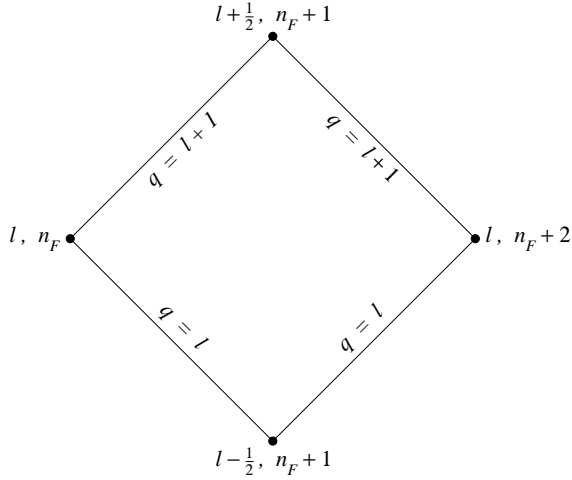}
\end{center}
\caption{The diamond structure of a supermultiplet, with supersymmetry
fractions $q$.}
\label{fig:Qfract}
\end{figure}

A supermultiplet can be constructed by considering $Q^\dagger_m$
as creation operators and $Q_m$ as annihilation operators acting
on the ``vacuum'' -- the lowest-$n_F$ state of the supermultiplet.
Starting from a single eigenstate of $H$ with nonzero energy
and definite $n_F\le3$, $j$, and $m$, Eq.\ (\ref{QQcom}) implies
that exactly three more states are produced: two by acting with
$Q^\dagger_{\half}$ and $Q^\dagger_{-\half}$, with quantum numbers
$n_F+1$ and $m\pm\half$, and one by acting with $Q^\dagger_{\half}
Q^\dagger_{-\half} - Q^\dagger_{-\half} Q^\dagger_{\half}$, with
quantum numbers $n_F+1$, $j$, and and $m$ (since
$Q^\dagger_{\half} Q^\dagger_{-\half} + Q^\dagger_{-\half}
Q^\dagger_{\half}$ vanishes). It is easy to show that applying
$Q_m$ or $Q^\dagger_m$ to any of these states either gives zero or
another of these states.  Starting now from a full rotational
multiplet, $Q^\dagger_{m'}|n_f,j,m\rangle$ is decomposed into two
multiplets with $n_F+1$ and $j\pm\half$, ($j-\half$ is absent if
$j=0$); additionally, we have one $n_F,j$ and one $n_F+2,j$
multiplet. This structure, which is nothing but a diamond of
Sect.\ \ref{sec:diamonds}, is shown in Fig.\ \ref{fig:Qfract}.

\subsection{Supersymmetry fractions}
\label{sec:frac}

Applying Eqs.\ (\ref{eq:redadj}) and (\ref{eq:redprod})
to Eq.\ (\ref{QQcom}), we obtain
\begin{eqnarray}
&&{\langle j;i''\Vert H\Vert j;i\rangle} = \nonumber \\
&&\quad{1\over4\sqrt{2j+1}}\sum_{j',i'}\left[
    {\langle j ;i''\Vert Q^\dagger\Vert j';i' \rangle} \,
    {\langle j ;i  \Vert Q^\dagger\Vert j';i' \rangle} +
    {\langle j';i' \Vert Q^\dagger\Vert j ;i''\rangle} \,
    {\langle j';i' \Vert Q^\dagger\Vert j ;i  \rangle} \right].
\label{eq:QHred}
\end{eqnarray}
When $|j;i\rangle$ is an eigenstate of $H$ with eigenvalue $E_{j,i}$
and $i''=i$, Eq.\ (\ref{eq:QHred}) reduces to
\begin{equation}
\sum_{j',i'}\left[
    \left|{\langle j ;i \Vert Q^\dagger\Vert j';i'\rangle}\right|^2 +
    \left|{\langle j';i'\Vert Q^\dagger\Vert j ;i \rangle}\right|^2 \right] =
4(2j+1) E_{j,i}.
\label{eq:QEred}
\end{equation}
It is therefore natural to define the ``supersymmetry fractions''
\[ q(j',i'|j,i) \equiv {1 \over 4 E_{j,i}}
   \left|{\langle j';i'\Vert Q^\dagger\Vert j ;i \rangle}\right|^2, \]
which, in the limit of exact supersymmetry, satisfy the sum rule
\begin{equation}
\sum_{j',i'}[q(j',i'|j,i) + q(j,i|j',i')] = 2j + 1.
\label{eq:Qfrac}
\end{equation}

Supersymmetry fractions allow an easy classification of states into
supermultiplets: $q$ is nonzero only between states belonging to the
same supermultiplet.  For discrete states, the fractions can be
computed explicitly, using Eq.\ (\ref{eq:QEred}) and remembering that
$\langle j';i' \Vert ({\tilde Q}, Q^\dagger)_1\Vert j;i\rangle$
vanishes on gauge-invariant states.  When no mixing occurs, the
resulting fractions are
\begin{eqnarray}
q(n_F{+}1,j{+}\half|n_F,j) &=& q(n_F{+}2,j|n_F{+}1,j{+}\half) = j+1,
\nonumber \\
q(n_F{+}1,j{-}\half|n_F,j) &=& q(n_F{+}2,j|n_F{+}1,j{-}\half) = j,
\label{eq:Qfrac1}
\end{eqnarray}
(they are also shown in Fig.~\ref{fig:Qfract});
they saturate Eq.\ (\ref{eq:Qfrac}).

\subsection{The spectrum - cutoff dependence and general properties}

Monitoring the cutoff dependence is crucial for at least two
reasons. First, it provides a model-independent information about
the errors induced by limiting the Hilbert space. Second, it
allows to distinguish between localized and non-localized states.
The latter feature is particularly useful in studying
supersymmetric gauge systems where continuous and discrete spectra
are known to coexist. It was shown in Ref.\ \cite{TW} that
eigenenergies of non-localized states drop slowly to zero with the
cutoff, while the discrete spectrum is characterized by rapid
convergence to the finite, ``infinite volume'' eigenvalues.

\begin{figure}[tbp]
\begin{center}
\psfig{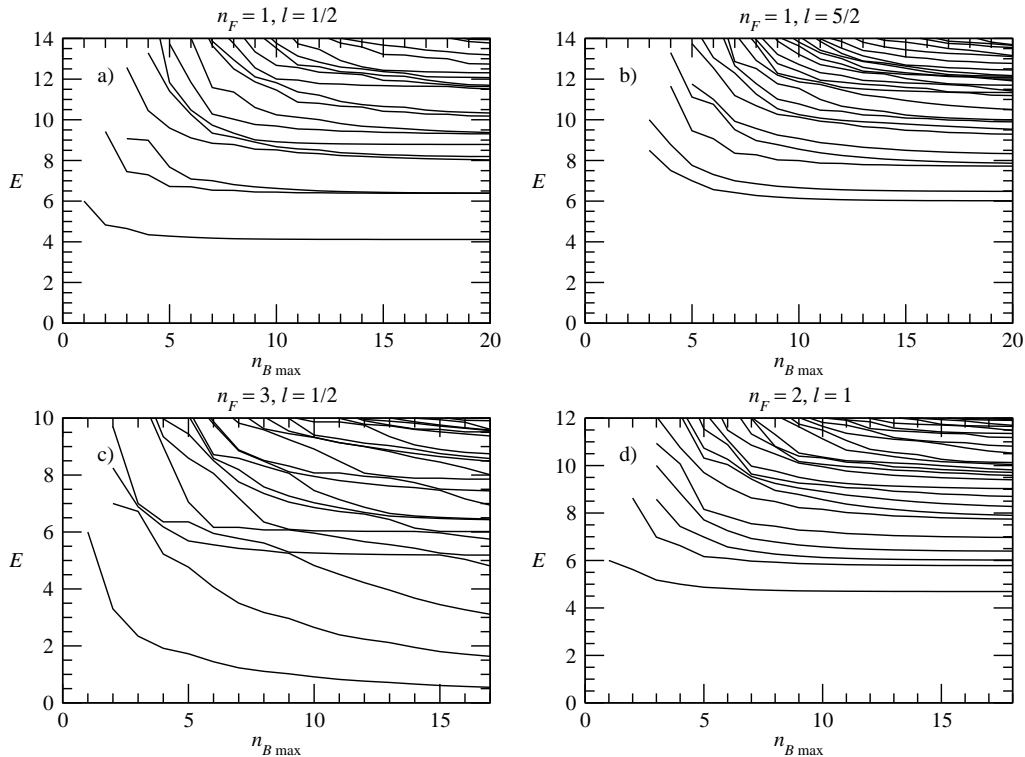}
\end{center}
\vskip-7mm
\caption{Cutoff dependence of the spectrum in a sample of channels.}
\label{fig:Ecombined}
\end{figure}

In the $(0,0)$ and $(2,0)$ channels our results are identical with
those of van Baal, Fig.\ \ref{fig:vanBaal}, hence we concentrate
on other channels, plotted in Fig.\ \ref{fig:Ecombined}. For
$n_F=1, j=\half$ the spectrum is similar to that in the $(0,0)$
channel. The levels are quickly converging and the available
cutoff is sufficient to guarantee small errors.  Similar situation
occurs for higher angular momenta with $n_F=1$, e.g., for
$j=\txf52$. Some degeneracies are observed, e.g., second and third
level of the channel $(n_F=1,j=\half)$. They are not caused by
supersymmetry, which connects states from different channels, as
was discussed in Sect.\ \ref{sec:SUSY-alg}.

On the other hand in the $n_F=3$ sector we clearly observe both
convergent, localized states and slowly falling ones from the
continuum.  The $j=\half$ channel plotted in Fig.\
\ref{fig:Ecombined}c is very similar in this respect to the
$n_F=2, j=0$ channel shown earlier (cf.\ Fig.\ \ref{fig:vanBaal}).
Similar behavior is seen for other angular momenta.  Again,
cutoffs reached with present method allow for quantitative studies
of many features of the localized states. Scattering states show
much more complexity, nevertheless some of their properties can be
also inferred, see below.

We therefore seem to confirm the general pattern suggested already by
the low \nbm\ calculations: in zero- and one-fermion sectors (and
their particle-hole images) the spectrum is discrete, while in the
``fermion rich'' sectors with $n_F=2,3,4$, both localized and
non-localized states coexist. There is however additional refinement
of this rule.

Contrary to earlier expectations the spectrum is entirely discrete
also in the $n_F=2, j=1$ channel, Fig.\ \ref{fig:Ecombined}d.
In fact we observe that this
happens in all $n_F=2$ channels with {\em odd} angular momentum $j$.
Therefore previous rule is modified to the following: scattering
states exist in the $n_F=3$ sector for all angular momenta, while
non-localized states with $n_F=2$ occur only for {\em even} angular
momentum. This will find yet simpler interpretation when we discuss in
detail the supermultiplet structure of the spectrum.

\subsection{Discrete spectrum - identifying supermultiplets}
\label{sec:identify}

\begin{figure}[tbp]
\begin{center}
\epsfig{width=10cm,height=8cm,file=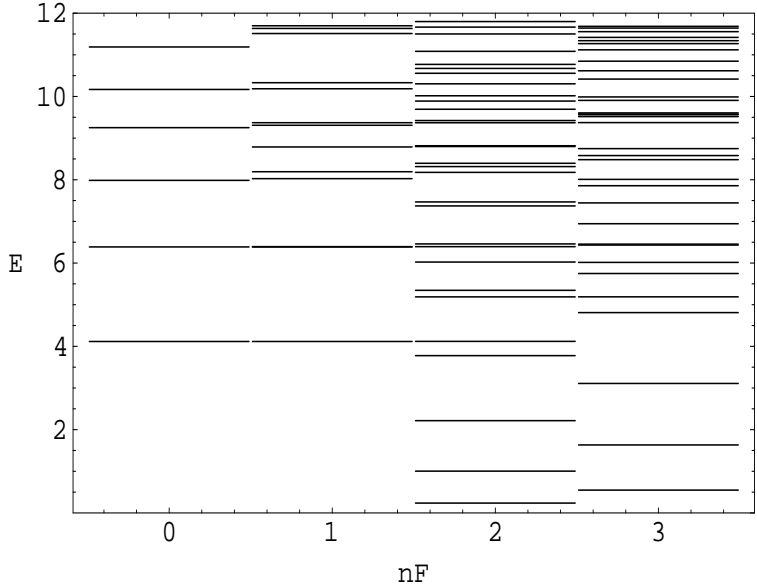}
\end{center}
\caption{Spectrum of the $j=0,\half$ states obtained with highest cutoff
available in each fermionic sector,
cf.\ Table \ref{tab:size-vs-nB}.}
\label{allFs}
\end{figure}

To begin with, let us collect the ``spectroscopy'' graph of the energy
levels with lowest angular momentum in all fermionic sectors,
Fig.\ \ref{allFs}.

Clearly a number of states in adjacent channels have identical
energies (within our cutoff errors) and are therefore good candidates
for SUSY partners. Confronting this with the cutoff dependence, Figs.\
\ref{fig:vanBaal}, \ref{fig:Ecombined}, we see that identification of
SUSY multiplets is simpler in the discrete part of the spectrum.
Restoration of supersymmetry among the non-localized states is more
complex and will be discussed later. Still, in order to achieve a
complete classification (even of localized states), it is important to
analyze together the highest cutoff results, the cutoff dependence,
and supersymmetry fractions. This is done below.

Recall from Sect.\ \ref{sec:SUSY-alg} that a supermultiplet of
SYMQM is composed by the diamond of O(3) multiplets shown in Fig.\
\ref{fig:Qfract}: $(n_F,j)$, the multiplet with the lowest $n_F$,
$(n_F{+}1,j{+}\half)$, $(n_F{+}1,j{-}\half)$ (only if $j>0$), and
$(n_F{+}2,j)$.  We will denote the full supermultiplet by the
spectroscopic labels $n_F(j)$
\footnote{Note that the pairs of supermultiplets conjugated by
particle-hole symmetry are $0(j)$ with $4(j)$ and $1(j)$ with $3(j)$,
while $2(j)$ is self-conjugated.}
for the ground state
in the channel,
$n_F(j)'$ for the first excited state etc.; when many excited states
are considered, we label them by their energies multiplied by $10^3$.
In the case $n_F=0$, when $(-1)^{n_B}$ is conserved for the $(0,j)$
multiplet, we add an $n_B$-parity label, i.e., we write $0(j^\pm)$.

\begin{figure}[tbp]
\begin{center}
\psfig{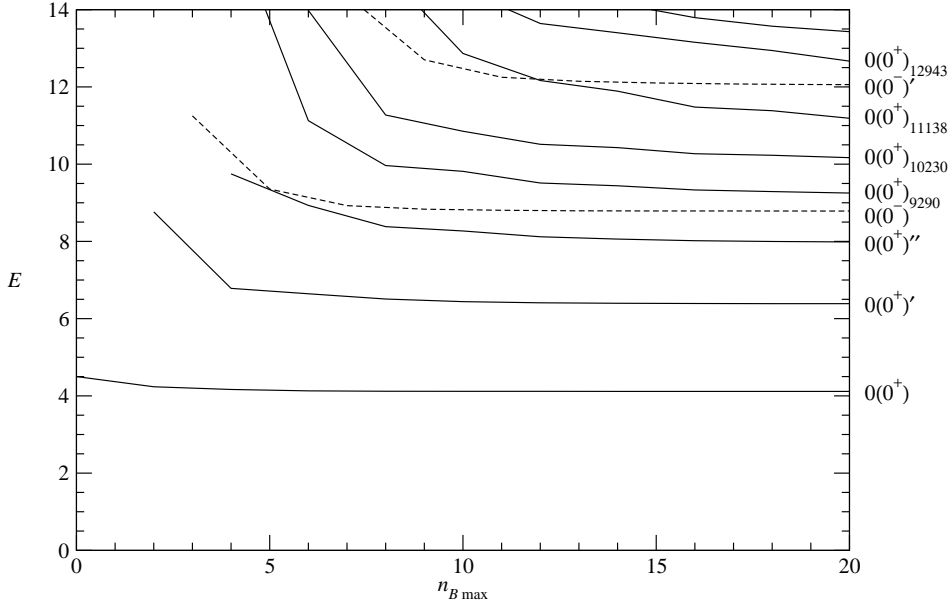}
\end{center}
\caption{Energy levels vs.\ $\nBmax$ for the channel $(0,0)$; solid
lines indicate even $n_B$-parity, dashed lines odd $n_B$-parity.}
\label{fig:E.F0.j0}
\end{figure}

\begin{figure}[tbp]
\begin{center}
\psfig{file=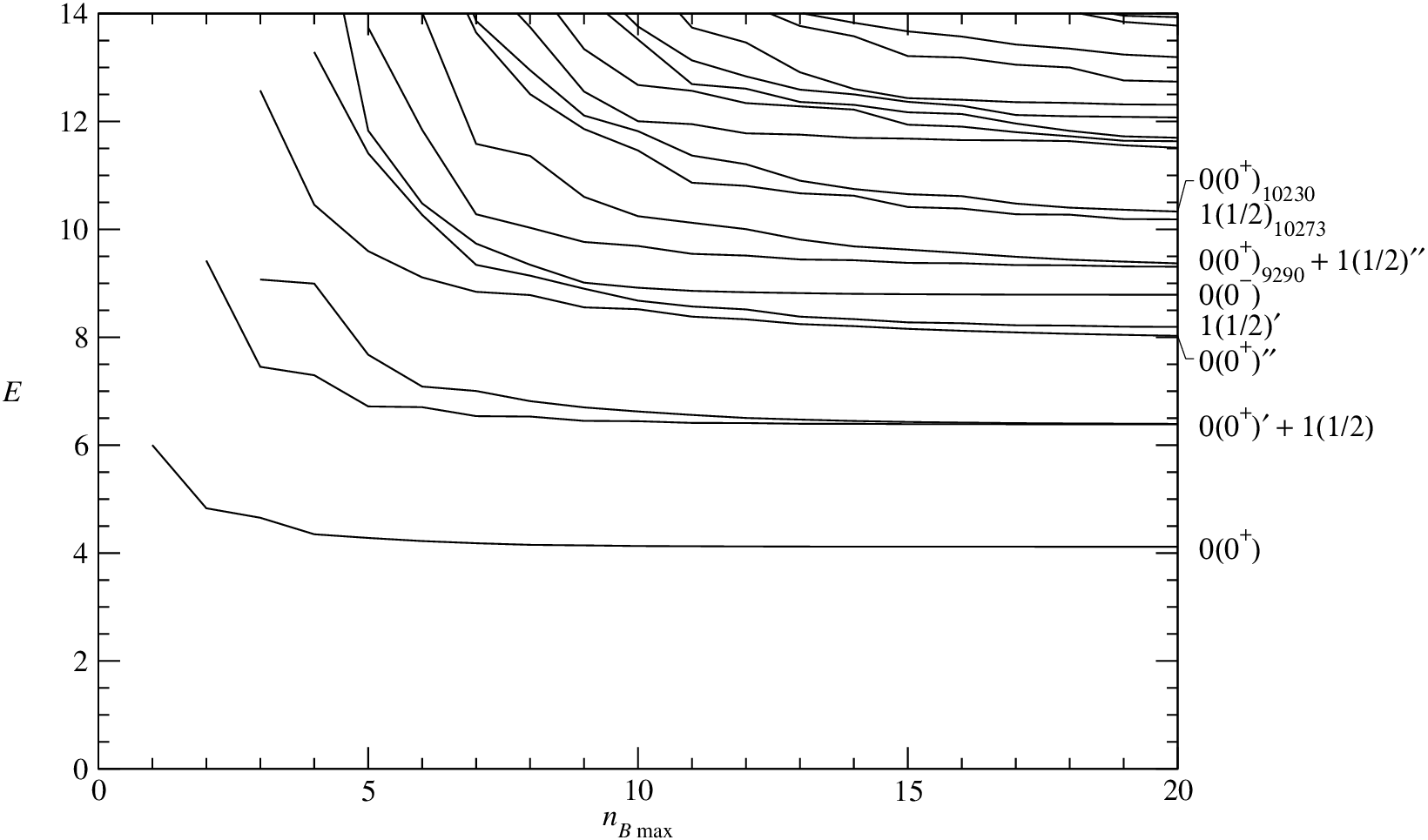,height=8cm}
\end{center}
\caption{Energy levels vs.\ $\nBmax$ for the channel $(1,\half)$.}
\label{fig:E.F1.j1h}
\end{figure}

\begin{figure}[tbp]
\begin{center}
\psfig{file=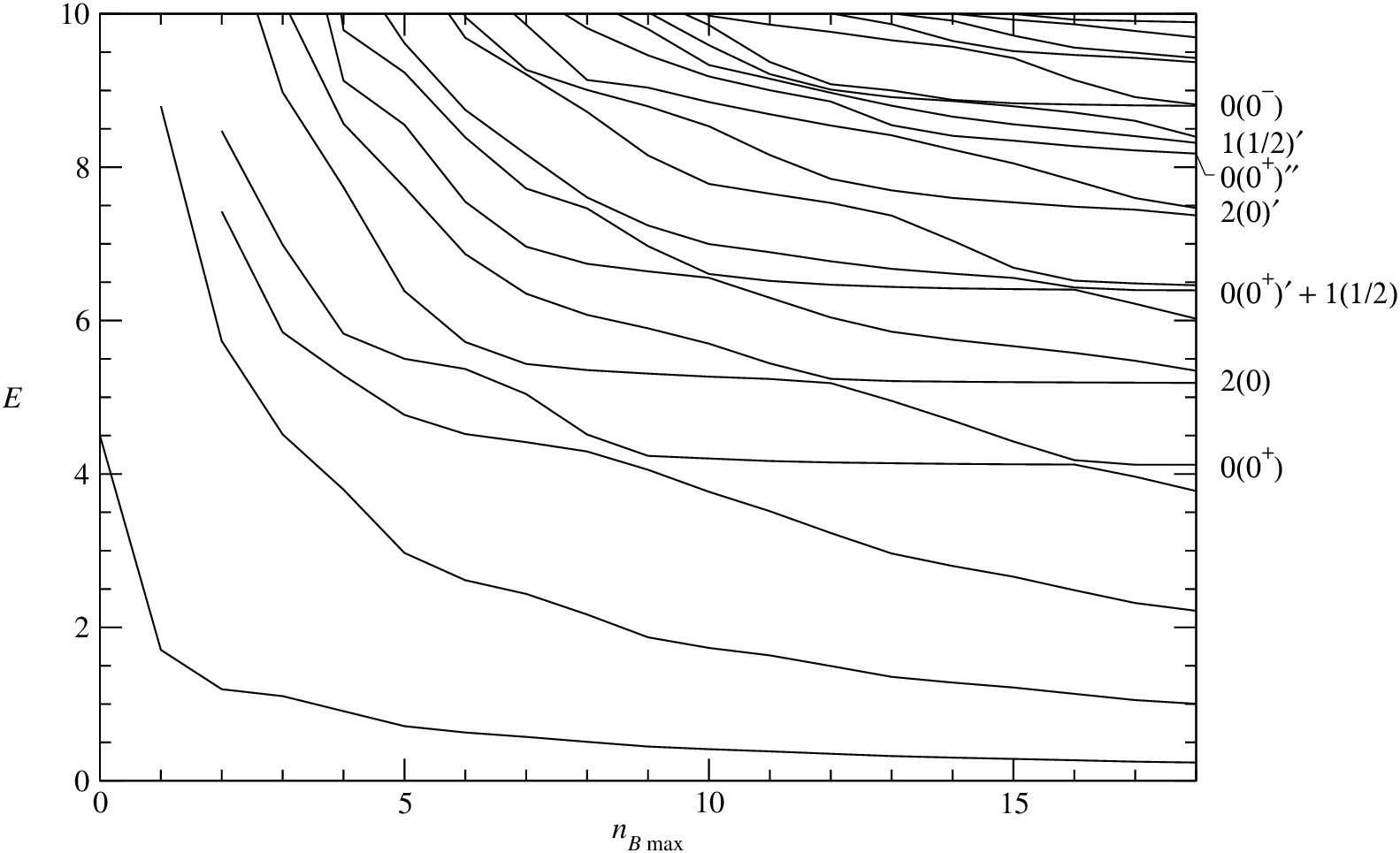,height=8cm}
\end{center}
\caption{Energy levels vs.\ $\nBmax$ for the channel $(2,0)$.}
\label{fig:E.F2.j0}
\end{figure}

\begin{figure}[tbp]
\begin{center}
\psfig{file=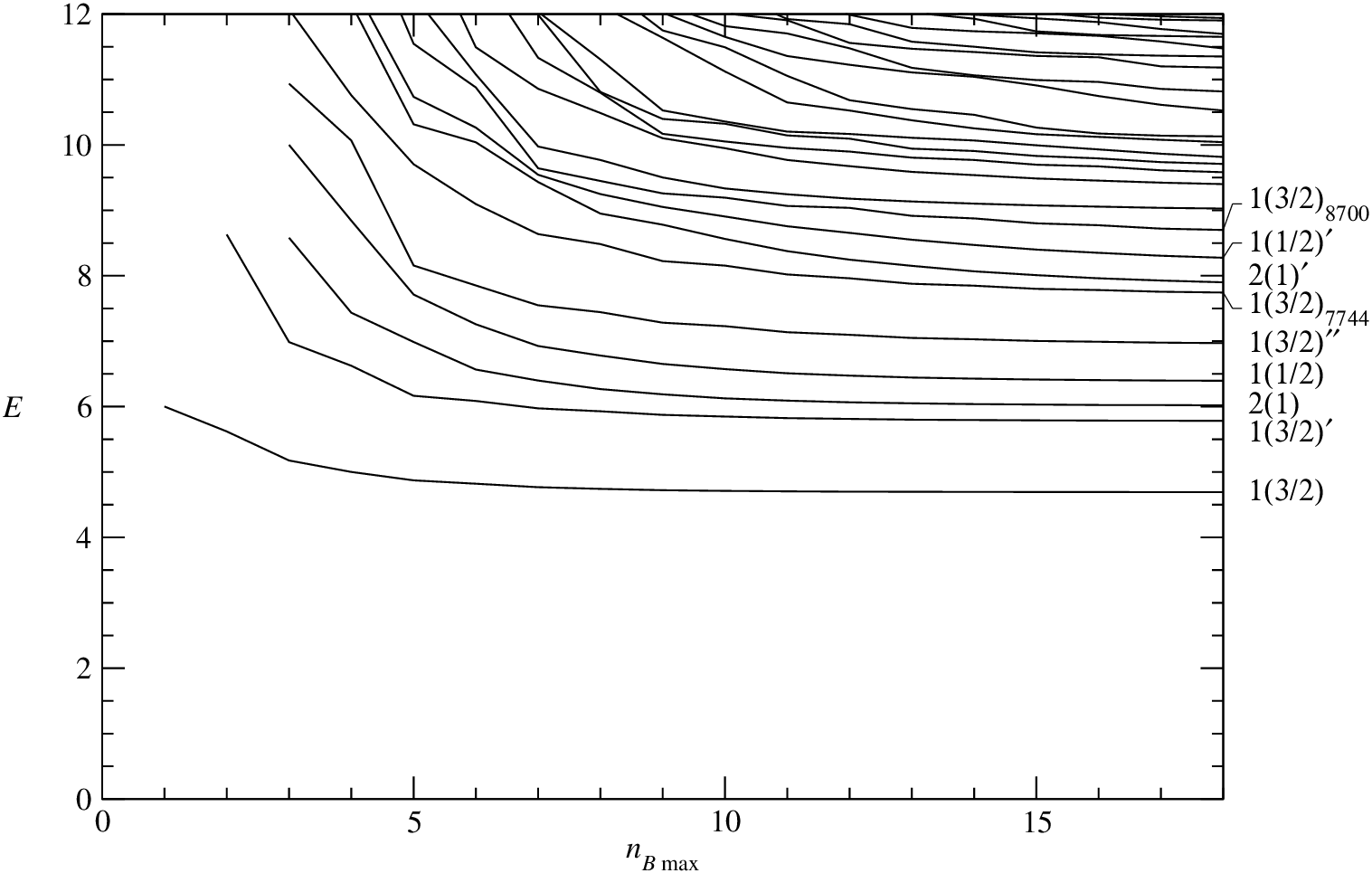,height=8cm}
\end{center}
\caption{Energy levels vs.\ $\nBmax$ for the channel $(2,1)$.}
\label{fig:E.F2.j1}
\end{figure}

\begin{figure}[tbp]
\begin{center}
\psfig{file=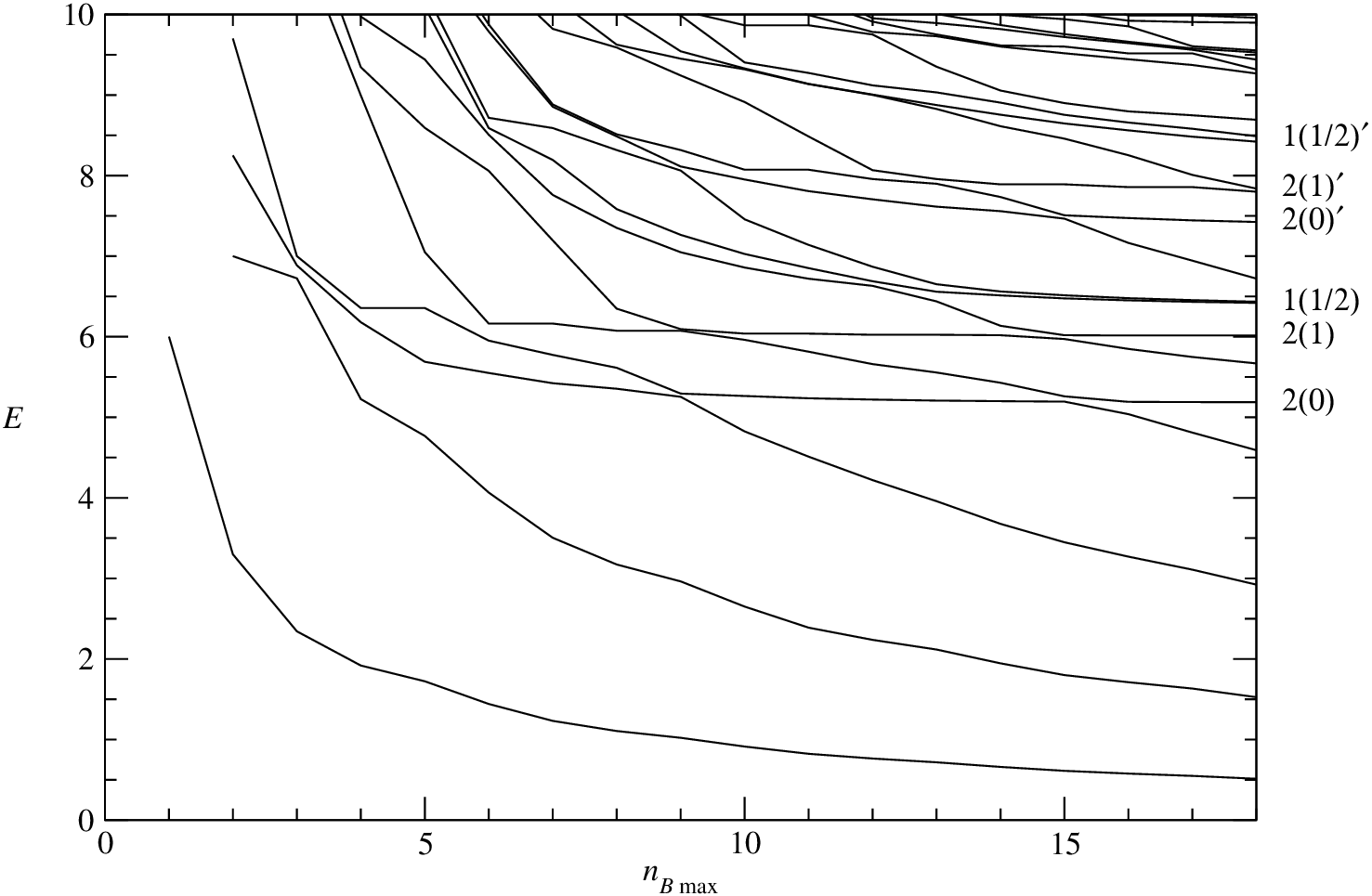,height=8cm}
\end{center}
\caption{Energy levels vs.\ $\nBmax$ for the channel $(3,{1\over2})$.}
\label{fig:E.F3.j1h}
\end{figure}

We begin the detailed presentation of our data by plotting the energy
levels vs.\ $\nBmax$ for several channels in Figs.\
\ref{fig:E.F0.j0}--\ref{fig:E.F3.j1h}.  The channels with higher $j$
follow a pattern quite similar to these, and therefore we will not
present here the corresponding plots; for the remaining channels with
$j\le4$, they can be found on the authors' web site \cite{ref:MCwww}.
For the lower levels of each channel, we quote the spectroscopic
labels $n_F(j)$ identifying the supermultiplet to which the level
belongs, anticipating results from the following of the present
Section.

The most effective tool to classify states into supermultiplets is
based on the analysis of supersymmetry fractions; the spectroscopic
labels reported in the above-mentioned plots are obtained by the
following method.

Let us select two sectors with fixed $(n_F'=n_F{+}1, j'=j\pm\half)$
and $(n_F, j)$, and construct the matrix $q(n_F',j',i'|n_F,j,i)$,
where $i'$ and $i$ run over the energy eigenvalues of the two sectors;
the cutoff $\nBmax$ is often the same in the two sector, but it may be
different, in which case we will write the two cutoffs as
$\nBmax'|\nBmax$.
Take one $(n_F',j',i')$ state and look at the corresponding row of the
matrix as $\nBmax$ grows: if all elements go to zero, the state
belongs to a $(n_F',j')$ supermultiplet.  In the same way, take one
$(n_F,j,i)$ state and look at the corresponding column of the matrix:
if all elements go to zero, the state belongs to a $(n_F{-}2,j)$
supermultiplet.
If one element remains nonzero, the two corresponding states belong to
the same supermultiplet: we look for the remaining superpartners
coupling to these two states in the appropriate channels, forming the
diamond of Fig.~\ref{fig:Qfract}, with the given values of $q$.
If two elements remain nonzero, we have a case of ``accidental''
degeneracy of two supermultiplets: the $q$'s are the superposition of
two patterns of Fig.~\ref{fig:Qfract}, with coefficients
$\cos^2\theta$ and $\sin^2\theta$, where $\theta$ is the mixing
angle between the energy eigenstates (which are not exactly degenerate
at finite $\nBmax$) and the states belonging to a definite
supermultiplet.
If a number of elements remain nonzero (typically 5 to 10 for our
values of $\nBmax$), the state belongs to the continuum.

\begin{figure}[tbp]
\begin{center}
\epsfig{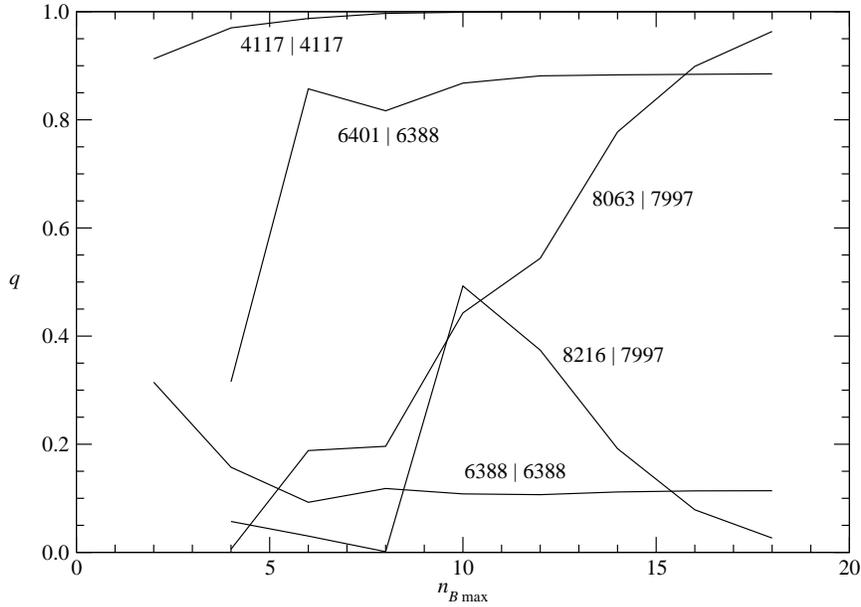}
\end{center}
\caption{Selected supersymmetry fractions $q(1,\half,i'|0,0^+,i)$
     vs.\ $\nBmax$.}
\label{fig:Qfrac1}
\end{figure}

Let us look in details, e.g., at the transition $q(1,\half|0,0^+)$.
The $q$ matrix for our highest value of $\nBmax$ is shown in Table
\ref{tab:Qfrac:11h:00} for $\nBmax=18$; selected coefficients are
plotted vs.\ $\nBmax$ in Fig.\ \ref{fig:Qfrac1}.  We identify the
states in each channels by their energies at $\nBmax=18$ ($\nBmax=19$
for $(0,j^-)$), multiplied by $10^3$: we use the notation $(n_F,j,E)$,
or just $E$ when $n_F$ and $j$ are obvious.

\begin{table}[tbp]
\begin{center}
\begin{tabular}{|r|rrrrrrrrrr|}
\multicolumn{1}{c}{} &
\multicolumn{10}{c}{$(0,0^+)$} \\
\hline
 $(1,1/2)$ & 4117 & 6388 & 7997 & 9290 & 10230 & 11383 & 12943 & 13572 & 14109 & 14955 \\
\hline
4117 & 1000 & 0 & 0 & 0 & 0 & 0 & 0 & 0 & 0 & 0 \\
6388 & 0 & 114 & 0 & 0 & 0 & 0 & 0 & 0 & 0 & 0 \\
6401 & 0 & 885 & 0 & 0 & 0 & 0 & 0 & 0 & 0 & 0 \\
8063 & 0 & 0 & 963 & 4 & 1 & 0 & 0 & 0 & 0 & 0 \\
8216 & 0 & 0 & 26 & 2 & 1 & 0 & 0 & 0 & 0 & 0 \\
8789 & 0 & 0 & 0 & 0 & 0 & 0 & 0 & 0 & 0 & 0 \\
9334 & 0 & 0 & 0 & 84 & 0 & 0 & 0 & 0 & 0 & 0 \\
9438 & 0 & 0 & 3 & 877 & 24 & 5 & 0 & 0 & 0 & 0 \\
10273 & 0 & 0 & 0 & 1 & 124 & 0 & 0 & 0 & 0 & 0 \\
10402 & 0 & 0 & 1 & 16 & 817 & 35 & 2 & 0 & 0 & 0 \\
11637 & 0 & 0 & 0 & 0 & 0 & 12 & 0 & 0 & 0 & 0 \\
11726 & 0 & 0 & 0 & 5 & 15 & 662 & 22 & 4 & 0 & 0 \\
11827 & 0 & 0 & 0 & 1 & 4 & 221 & 30 & 9 & 0 & 1 \\
12097 & 0 & 0 & 0 & 0 & 0 & 1 & 0 & 0 & 0 & 0 \\
12344 & 0 & 0 & 0 & 0 & 0 & 0 & 1 & 0 & 0 & 0 \\
13000 & 0 & 0 & 0 & 0 & 1 & 4 & 54 & 1 & 0 & 1 \\
13350 & 0 & 0 & 0 & 0 & 1 & 10 & 611 & 289 & 3 & 18 \\
14037 & 0 & 0 & 0 & 1 & 2 & 15 & 132 & 351 & 19 & 72 \\
14140 & 0 & 0 & 0 & 0 & 0 & 2 & 18 & 44 & 777 & 16 \\
14190 & 0 & 0 & 0 & 0 & 0 & 1 & 8 & 49 & 5 & 5 \\
\hline
\end{tabular}
\end{center}
\caption{Matrix of supersymmetry fractions $q(1,\half|0,0^+)$ at
    $\nBmax=18$.  States are identified by their energies.  All
    numbers are multipled by $10^3$.}
\label{tab:Qfrac:11h:00}
\end{table}

Proceeding by increasing energies, first we see a perfect match for
the two ground states, i.e., $q(4117|4117)=1$, and they therefore
belong to the $0(0^+)$ supermultiplet.  Next we have a case of mixing:
since $q(6388|6388)=0.114$ and $q(6401|6388)=0.885$, $(0,0^+,6388)$
belongs to $0(0^+)'$, while $(1,\half,6388)$ and $(1,\half,6401)$ are
linear combination of states belonging to the ``accidentally''
degenerate $0(0^+)'$ and $(1,\half)$ supermultiplets, with a mixing
angle $\theta$ with $\cos^2\theta=0.114$.
For higher levels, $q$'s are not completely stable in $\nBmax$, and we
need to extrapolate them to $\nBmax\to\infty$.  We clearly see that
$(0,0^+,7997)$ and $(1,\half,8063)$ belong to $0(0^+)''$.  For higher
states, the analysis requires more care.

The remaining members of the $0(0^+)$ supermultiplets can be
identified by looking at $q(2,0,i'|1,\half,i)$ , which is shown in
Table \ref{tab:Qfrac:20:11h} for $\nBmax=17|18$: $(2,0,4121)$ belongs
to $0(0^+)$; since $q(6397|6404)\to0$ and $q(6484|6388)\to0$,
$(2,0,6397)$ and $(2,0,6404)$ are linear combination of states
belonging $0(0^+)'$ and $(1,\half)$, with the same mixing angle
$\theta$ as above.  $(2,0,8806)$, despite the high energy, is very
easily attributed to $0(0^-)$, with the help of Table
\ref{tab:Qfrac:11h:00m}.  The levels $(2,0,i')$ related to the
continuum spectrum have zero $q(2,0,i'|1,\half,i)$.

\begin{table}[tbp]
\begin{center}
\begin{tabular}{|r|rrrrrrrr|}
\multicolumn{1}{c}{} &
\multicolumn{8}{c}{$(1,1/2)$} \\
\hline
 $(2,0)$ & 4117 & 6388 & 6401 & 8063 & 8216 & 8789 & 9334 & 9438 \\
\hline
237 & 0 & 0 & 0 & 0 & 0 & 0 & 0 & 0 \\
1004 & 0 & 0 & 0 & 0 & 0 & 0 & 0 & 0 \\
2216 & 0 & 0 & 0 & 0 & 0 & 0 & 0 & 0 \\
3777 & 0 & 0 & 0 & 0 & 0 & 0 & 0 & 0 \\
4119 & 1000 & 0 & 0 & 0 & 0 & 0 & 0 & 0 \\
5188 & 0 & 0 & 0 & 0 & 0 & 0 & 0 & 0 \\
5345 & 0 & 0 & 0 & 0 & 0 & 0 & 0 & 0 \\
6024 & 0 & 0 & 0 & 0 & 0 & 0 & 0 & 0 \\
6394 & 0 & 556 & 34 & 0 & 0 & 0 & 0 & 0 \\
6459 & 0 & 1 & 903 & 3 & 0 & 0 & 0 & 0 \\
7372 & 0 & 0 & 0 & 0 & 0 & 0 & 0 & 0 \\
7468 & 0 & 0 & 0 & 2 & 1 & 0 & 0 & 0 \\
8177 & 0 & 0 & 1 & 667 & 82 & 0 & 1 & 3 \\
8317 & 0 & 0 & 1 & 253 & 192 & 0 & 2 & 14 \\
8396 & 0 & 0 & 0 & 29 & 228 & 0 & 1 & 8 \\
8798 & 0 & 0 & 0 & 0 & 0 & 999 & 0 & 0 \\
8817 & 0 & 0 & 0 & 1 & 0 & 0 & 2 & 0 \\
9369 & 0 & 0 & 0 & 0 & 0 & 0 & 165 & 7 \\
9422 & 0 & 0 & 0 & 1 & 1 & 0 & 352 & 18 \\
9693 & 0 & 0 & 0 & 8 & 3 & 0 & 7 & 819 \\
\hline
\end{tabular}
\end{center}
\caption{Matrix of supersymmetry fractions $q(2,0|1,\half)$ at
    $\nBmax=18$.}
\label{tab:Qfrac:20:11h}
\end{table}

\begin{table}[tbp]
\begin{center}
\begin{tabular}{|r|rrr|}
\multicolumn{1}{c}{} &
\multicolumn{3}{c}{$(0,0^-)$} \\
\hline
 $(1,1/2)$ & 8787 & 12063 & 14064 \\
\hline
4117 & 0 & 0 & 0 \\
6388 & 0 & 0 & 0 \\
6401 & 0 & 0 & 0 \\
8063 & 0 & 0 & 0 \\
8216 & 0 & 0 & 0 \\
8789 & 1000 & 0 & 0 \\
9334 & 0 & 0 & 0 \\
9438 & 0 & 0 & 0 \\
10273 & 0 & 0 & 0 \\
10402 & 0 & 0 & 0 \\
11637 & 0 & 0 & 0 \\
11726 & 0 & 1 & 0 \\
11827 & 0 & 0 & 0 \\
12097 & 0 & 990 & 1 \\
12344 & 0 & 5 & 0 \\
\hline
\end{tabular}
\end{center}
\caption{Matrix of supersymmetry fractions $q(1,{1\over2}|0,0^-)$
         at $\nBmax=18|19$.}
\label{tab:Qfrac:11h:00m}
\end{table}

We then look at $q(2,1,i'|1,\half,i)$ and $q(3,\half,i'|2,0,i)$ to
identify the remaining members of the $1(\half)$ supermultiplets.
$q(2,1,i'|1,\half,i)$, presented in Table \ref{tab:Qfrac:21:11h},
presents the same pattern as $q(2,0,i'|1,\half,i)$, except for the
absence of continuum states, and we will not delve into the
classification of states.

\begin{table}[tbp]
\begin{center}
\begin{tabular}{|r|rrrrrrrrrrr|}
\multicolumn{1}{c}{} &
\multicolumn{11}{c}{$(1,1/2)$} \\
\hline
 $(2,1)$ & 4117 & 6388 & 6401 & 8063 & 8216 & 8789 & 9334 & 9438 & 10273 & 10402 & 11637 \\
\hline
4692 & 0 & 0 & 0 & 0 & 0 & 0 & 0 & 0 & 0 & 0 & 0 \\
5783 & 0 & 0 & 0 & 0 & 0 & 0 & 0 & 0 & 0 & 0 & 0 \\
6019 & 0 & 0 & 0 & 0 & 0 & 0 & 0 & 0 & 0 & 0 & 0 \\
6395 & 0 & 1328 & 171 & 0 & 0 & 0 & 0 & 0 & 0 & 0 & 0 \\
6971 & 0 & 0 & 0 & 0 & 0 & 0 & 0 & 0 & 0 & 0 & 0 \\
7744 & 0 & 0 & 0 & 0 & 0 & 0 & 0 & 0 & 0 & 0 & 0 \\
7899 & 0 & 0 & 0 & 0 & 0 & 0 & 0 & 0 & 0 & 0 & 0 \\
8275 & 0 & 0 & 0 & 42 & 1440 & 0 & 6 & 0 & 4 & 0 & 0 \\
8700 & 0 & 0 & 0 & 0 & 0 & 0 & 0 & 0 & 0 & 0 & 0 \\
9027 & 0 & 0 & 0 & 0 & 0 & 0 & 0 & 0 & 0 & 0 & 0 \\
9400 & 0 & 0 & 0 & 0 & 3 & 0 & 1341 & 124 & 22 & 0 & 0 \\
9583 & 0 & 0 & 0 & 0 & 0 & 0 & 0 & 0 & 0 & 0 & 0 \\
9710 & 0 & 0 & 0 & 0 & 0 & 0 & 0 & 0 & 0 & 0 & 0 \\
9817 & 0 & 0 & 0 & 0 & 0 & 0 & 0 & 0 & 0 & 0 & 0 \\
10042 & 0 & 0 & 0 & 0 & 0 & 0 & 0 & 0 & 0 & 0 & 0 \\
10129 & 0 & 0 & 0 & 0 & 0 & 0 & 0 & 0 & 0 & 0 & 0 \\
10526 & 0 & 0 & 0 & 0 & 2 & 0 & 16 & 0 & 1231 & 167 & 3 \\
10817 & 0 & 0 & 0 & 0 & 0 & 0 & 0 & 0 & 0 & 0 & 0 \\
11183 & 0 & 0 & 0 & 0 & 0 & 0 & 0 & 0 & 0 & 0 & 0 \\
11352 & 0 & 0 & 0 & 0 & 0 & 0 & 0 & 0 & 0 & 0 & 0 \\
11475 & 0 & 0 & 0 & 0 & 0 & 0 & 0 & 0 & 0 & 0 & 0 \\
11652 & 0 & 0 & 0 & 0 & 0 & 0 & 0 & 0 & 0 & 0 & 1446 \\
11696 & 0 & 0 & 0 & 0 & 0 & 0 & 0 & 0 & 0 & 0 & 0 \\
\hline
\end{tabular}
\end{center}
\caption{Matrix of supersymmetry fractions $q(2,1|1,\half)$ at
    $\nBmax=18$.}
\label{tab:Qfrac:21:11h}
\end{table}

$q(3,\half,i'|2,0,i)$, shown in Table \ref{tab:Qfrac:31h:20}, presents
a new, very interesting pattern: we see states with a broad
distribution of $q$'s quite different from zero, even with states with
very different energies; looking at the $\nBmax$ dependence of the
levels, cf.\ Figs.\ \ref{fig:E.F2.j0} and \ref{fig:E.F3.j1h}, we
conclude that the patterns identifies continuum levels.  On the other
hand, $q$ is zero between continuum and discrete states, or between
discrete states of significantly different energies.  We can easily
identify members of supermultiplets with quantum numbers $0(0)$,
$1(\half)$, $2(0)$, and $2(1)$, with the remaining states belonging to
the continuum.  The ``doubling'' of $(3,\half)$ states belonging to
$1(\half)$ supermultiplets is due to the particle-hole symmetry, as
will be explained below.

\begin{table}[tbp]
\begin{center}
\begin{tabular}{|r|rrrrrrrrrrrr|}
\multicolumn{1}{c}{} &
\multicolumn{12}{c}{$(2,0)$} \\
\hline
 $(3,1/2)$ & 237 & 1004 & 2216 & 3777 & 4119 & 5188 & 5345 & 6024 & 6394 & 6459 & 7372 & 7468 \\
\hline
513 & 737 & 220 & 11 & 1 & 0 & 0 & 0 & 0 & 0 & 0 & 0 & 0 \\
1526 & 73 & 538 & 280 & 15 & 0 & 0 & 1 & 0 & 0 & 0 & 0 & 0 \\
2924 & 12 & 43 & 476 & 301 & 0 & 0 & 13 & 6 & 0 & 0 & 1 & 0 \\
4592 & 1 & 6 & 34 & 449 & 0 & 0 & 275 & 68 & 0 & 0 & 8 & 2 \\
5187 & 0 & 0 & 0 & 0 & 0 & 999 & 0 & 0 & 0 & 0 & 0 & 0 \\
5669 & 0 & 0 & 3 & 21 & 0 & 0 & 500 & 414 & 0 & 0 & 10 & 2 \\
6015 & 0 & 0 & 0 & 0 & 0 & 0 & 0 & 0 & 0 & 0 & 0 & 0 \\
6419 & 0 & 0 & 0 & 0 & 0 & 0 & 0 & 0 & 204 & 44 & 0 & 0 \\
6434 & 0 & 0 & 0 & 0 & 0 & 0 & 1 & 2 & 203 & 43 & 2 & 1 \\
6721 & 0 & 1 & 4 & 22 & 0 & 0 & 86 & 359 & 2 & 1 & 246 & 84 \\
7425 & 1 & 0 & 0 & 0 & 0 & 0 & 0 & 0 & 0 & 0 & 219 & 750 \\
7800 & 0 & 0 & 0 & 4 & 0 & 0 & 12 & 30 & 0 & 0 & 385 & 71 \\
7839 & 0 & 0 & 0 & 0 & 0 & 0 & 0 & 0 & 0 & 0 & 0 & 0 \\
8422 & 0 & 0 & 0 & 0 & 0 & 0 & 0 & 0 & 0 & 0 & 1 & 3 \\
\hline
\end{tabular}
\end{center}
\caption{Matrix of supersymmetry fractions $q(3,\txf12|2,0)$
         at $\nBmax=18$.}
\label{tab:Qfrac:31h:20}
\end{table}

It is also worth presenting $q(3,\half,i'|2,1,i)$, shown in Table
\ref{tab:Qfrac:31h:21}; thanks to the absence of continuum states from
$(2,1)$, it is very easy to identify states in $(3,\half)$ belonging
to the supermultiplets $1(\half)$ and $2(1)$.

\begin{table}[tbp]
\begin{center}
\begin{tabular}{|r|rrrrrrrrrrrr|}
\multicolumn{1}{c}{} &
\multicolumn{12}{c}{$(2,1)$} \\
\hline
 $(3,1/2)$ & 4692 & 5783 & 6019 & 6395 & 6971 & 7744 & 7899 & 8275 & 8700 & 9027 & 9400 & 9583 \\
\hline
513 & 0 & 0 & 0 & 0 & 0 & 0 & 0 & 0 & 0 & 0 & 0 & 0 \\
1526 & 0 & 0 & 0 & 0 & 0 & 0 & 0 & 0 & 0 & 0 & 0 & 0 \\
2924 & 0 & 0 & 0 & 0 & 0 & 0 & 0 & 0 & 0 & 0 & 0 & 0 \\
4592 & 0 & 0 & 0 & 0 & 0 & 0 & 0 & 0 & 0 & 0 & 0 & 0 \\
5187 & 0 & 0 & 0 & 0 & 0 & 0 & 0 & 0 & 0 & 0 & 0 & 0 \\
5669 & 0 & 0 & 0 & 0 & 0 & 0 & 0 & 0 & 0 & 0 & 0 & 0 \\
6015 & 0 & 0 & 999 & 0 & 0 & 0 & 0 & 0 & 0 & 0 & 0 & 0 \\
6419 & 0 & 0 & 0 & 748 & 0 & 0 & 0 & 1 & 0 & 0 & 0 & 0 \\
6434 & 0 & 0 & 0 & 744 & 0 & 0 & 0 & 1 & 0 & 0 & 0 & 0 \\
6721 & 0 & 0 & 0 & 3 & 0 & 0 & 0 & 1 & 0 & 0 & 0 & 0 \\
7425 & 0 & 0 & 0 & 0 & 0 & 0 & 0 & 5 & 0 & 0 & 0 & 0 \\
7800 & 0 & 0 & 0 & 0 & 0 & 0 & 0 & 8 & 0 & 0 & 0 & 0 \\
7839 & 0 & 0 & 0 & 0 & 0 & 21 & 964 & 0 & 1 & 0 & 0 & 1 \\
8422 & 0 & 0 & 0 & 0 & 0 & 0 & 0 & 721 & 0 & 0 & 8 & 0 \\
8489 & 0 & 0 & 0 & 1 & 0 & 0 & 0 & 641 & 0 & 0 & 16 & 0 \\
8693 & 0 & 0 & 0 & 0 & 0 & 0 & 0 & 60 & 0 & 0 & 15 & 0 \\
9267 & 0 & 0 & 0 & 0 & 0 & 0 & 0 & 0 & 0 & 0 & 2 & 0 \\
9317 & 0 & 0 & 0 & 0 & 0 & 0 & 0 & 1 & 0 & 0 & 125 & 0 \\
9441 & 0 & 0 & 0 & 0 & 0 & 0 & 4 & 0 & 4 & 3 & 0 & 678 \\
9529 & 0 & 0 & 0 & 0 & 0 & 0 & 0 & 8 & 0 & 0 & 594 & 0 \\
9554 & 0 & 0 & 0 & 0 & 0 & 0 & 0 & 15 & 0 & 0 & 682 & 0 \\
9898 & 0 & 0 & 0 & 0 & 0 & 0 & 0 & 0 & 0 & 0 & 1 & 0 \\
\hline
\end{tabular}
\end{center}
\caption{Matrix of supersymmetry fractions $q(3,\txf12|2,1)$
         at $\nBmax=18$.}
\label{tab:Qfrac:31h:21}
\end{table}

The analysis of $q$ for higher $j$ is repeated exactly in the same
way.  We will not present here the corresponding $q$ matrices, which
can be found in Ref.\ \cite{ref:MCwww} for the remaining channels with
$j\le4$.  We only remark that all $q(n_F',j'|n_F,j)$ for the same
values of $n_F',n_F$ and different values of $j',j$ are qualitatively
very similar (in the case of $n_F'=2$ ($n_F=2$), only for $j'$ ($j$)
having the same parity).

{}From all the above data, we can compile the spectroscopy of Tables
\ref{tab:spectrum} and \ref{tab:spectrum2}.  The table is limited to
$n_F\le2$, since the other supermultiplets can be obtained by
particle-hole reflection, and to $j\le4$, since nothing new happens
for higher $j$.

One feature should be stressed: for each $1(j)$ supermultiplet, the
particle-hole symmetry implies the existence of a conjugate
supermultiplet $3(j)$, and therefore of two $(3,j)$ states of
degenerate energy (in the $\nBmax\to\infty$ limit); we observe mixing
of each pair, with a mixing angle $\theta=\pi/4$.

\begin{figure}[tbp]
\begin{center}
\epsfig{width=10cm,file=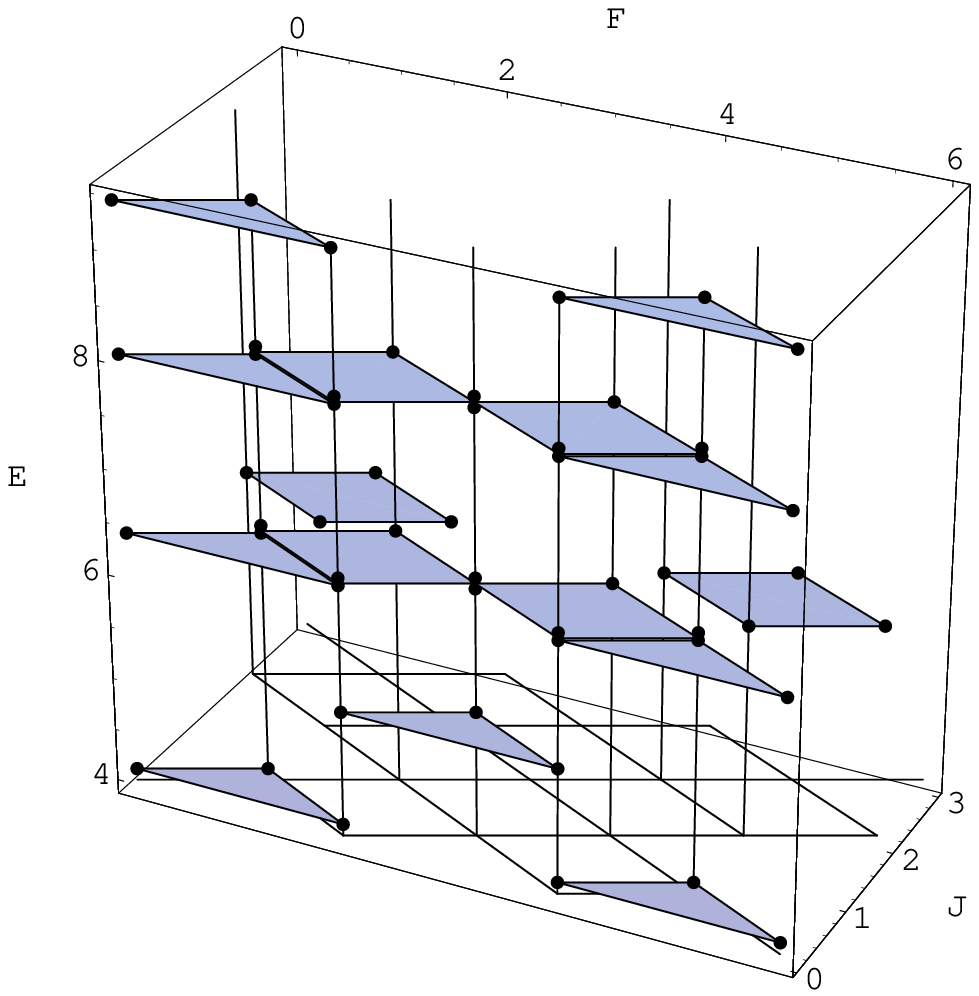}
\end{center}
\caption{Spectrum of a sample of supermultiplets identified in Tables
\ref{tab:spectrum} and \ref{tab:spectrum2}}
\label{fig3D}
\end{figure}

Figure \ref{fig3D} shows a sample of the lowest supermultiplets
for the first few angular momenta and all $n_F$. Degenerate
supermultiplets at $E\sim 6.4$ and $8.1$ were slightly split for
the sake of illustration.

\begin{table}[tbp]
\begin{center}
\begin{tabular}{|l|c|c|c|c|}
\hline
\multicolumn{5}{|c|}{$10^3{\times}$energies at $\nBmax=18$} \\
\hline
$n_F(j)$&$(n_F,j)$&\kern-5pt$(n_F{+}1,j{-}\half)$\kern-5pt&\
\kern-5pt$(n_F{+}1,j{+}\half)$\kern-5pt&\kern-5pt$(n_F{+}2,j)$\kern-5pt\\
\hline
$0(0^+)$  & \04117  &   ---    & \04117   &  \04119    \\
$0(0^+)'$ & \06388  &   ---    &6388, 6401& 6394, 6459 \\
$0(0^+)''$& \07997  &   ---    & \08063   &  \08177    \\
$0(0^+)_{9290}$& \09290 & ---  &9334, 9438& \\
$0(0^+)_{10230}$& 10230 & ---  &  10402   & \\
$0(0^+)_{11383}$& 11383 & ---  &  11726   & \\
$0(0^+)_{12943}$& 12943 & ---  & & \\
\hline
$0(0^-)$  & \08787  &   ---    & \08789   &  \08798    \\
$0(0^-)'$ &  12063  &   ---    &  12097   &   \0       \\
\hline
$0(2^+)$  & \06015  & \06019   & \06020   &  \06041    \\
$0(2^+)'$ & \07839  & \07899   & \07902   &  \08071    \\
$0(2^+)''$& \09441  & \0   & \09628   &  \0    \\
$0(2^+)_{9961}$& \09961  & \0   & \0   &  \0    \\
$0(2^+)_{11183}$& 11183  & \0   & \0   &  \0    \\
$0(2^+)_{12096}$& 12096  & \0   & \0   &  \0    \\
$0(2^+)_{13005}$& 13005  & \0   & \0   &  \0    \\
\hline
$0(2^-)$  &  11334  &  11352   &  11355   &   11407    \\
$0(2^-)'$ &  14045  &  14131   &  14187   &       \\
\hline
$0(3^+)$  &  12138  &  12174   &  12178   &   12230    \\
$0(3^+)'$ &  15068  &    &    &      \\
$0(3^+)''$ & 17647  &    &    &      \\
$0(3^+)_{19294}$ & 19294  &    &    &      \\
\hline
$0(3^-)$  &  18140  &  18395   &  18663   &   \0       \\
\hline
$0(4^+)$  & \07739  & \07768   & \07772   &  \07863    \\
$0(4^+)'$ & \09411  & \09603 & \09604 & \\
$0(4^+)''$ & 11153  & & & \\
$0(4^+)_{12603}$  & 12603  & & & \\
$0(4^+)_{13152}$  & 13152  & & & \\
$0(4^+)_{14364}$  & 14364  & & & \\
\hline
$0(4^-)$  &  13747  &  13824   &  13841   &   13948    \\
\hline
\end{tabular}
\end{center}
\caption{Spectroscopy of SYMQM.}
\label{tab:spectrum}
\end{table}

\begin{table}[tbp]
\begin{center}
\begin{tabular}{|l|c|c|c|c|}
\hline
\multicolumn{5}{|c|}{$10^3{\times}$energies at $\nBmax=18$} \\
\hline
$n_F(j)$&$(n_F,j)$&\kern-5pt$(n_F{+}1,j{-}\half)$\kern-5pt&\
\kern-5pt$(n_F{+}1,j{+}\half)$\kern-5pt&\kern-5pt$(n_F{+}2,j)$\kern-5pt\\
\hline
$1(1/2)$ &6388, 6401&6394, 6459&\06395    & 6419, 6434 \\
$1(1/2)'$ & \08216  & \08317   &\08275    & 8422, 8489 \\
$1(1/2)''$ &9334, 9438& & & \\
$1(1/2)_{10273}$& 10273 & & & \\
\hline
$1(3/2)$  & \04692  & \04692   & \04694   & 4694, 4696 \\
$1(3/2)'$ & \05783  & \05783   & \05791   & 5790, 5799 \\
$1(3/2)''$& \06971  & \06971   & \07008   & 7010, 7047 \\
$1(3/2)_{7744}$& \07744  & \07744 & \07809 & 7826, 7839 \\
$1(3/2)_{8700}$& \08700  & \08700 & \08782 & 8831, 8908 \\
\hline
$1(5/2)$  & \06486  & \06484   & \06501   & 6488, 6508 \\
$1(5/2)'$ & \07733  & \07751   & \07763   & 7808, 7866 \\
$1(5/2)''$& \08379  & \08386 & 8420, 8527$^*$& 8411, 8485 \\
$1(5/2)_{9352}$& \09352 & & & \\
\hline
$1(7/2)$  & \06591  & \06593   & \06611   & 6613, 6628 \\
$1(7/2)'$ & \07515  & \07518   & \07553   & 7546, 7578 \\
$1(7/2)''$& \08428  & 8420, 8527$^*$& \08548 & 8579, 8654 \\
$1(7/2)_{9314}$& \09314  & & & \\
\hline
$2(0)$    & \05188  &   ---    & \05187   &  \05188    \\
$2(0)'$   & \07373  &   ---    & \07444   &  \07425    \\
\hline
$2(1)$    & \06019  & \06015   & \06028   &  \06019    \\
$2(1)'$   & \07899  & \07839   & \08006   &  \07899    \\
\hline
$2(2)$    & \06734  & \06722   & \06736   &  \06734    \\
\hline
$2(3)$    & \06085  & \06079   & \06093   &  \06085    \\
$2(3)'$   & \07826  & \07800   & \07876   &  \07826    \\
$2(3)''$  & \08208  & \08168   & \08304   &  \08208    \\
\hline
$2(4)$    & \08169  & \08140   & \0       &  \08169    \\
\hline
\end{tabular}
\end{center}
\caption{Spectroscopy of SYMQM (continued). \protect\\
$^*$ \footnotesize The two states $(2,3,8420)$ and $(2,3,8527)$ belong to
supermultiplets with different energies, but appear to be mixed at the
available values of $\nBmax$.}
\label{tab:spectrum2}
\end{table}

\subsection{Continuous spectrum}

We already mentioned that a continuous spectrum is observed for
all $j$'s in the $n_F=3$ channels, but only for even angular
momenta in the $n_F=2,4$ sectors.  This pattern is consistent with
supersymmetry, and simply means that continuous states exist only
in supermultiplets $2(j)$ with {\em even\/} $j$.  Note that the
``opposite'' behaviour (all $n_F=2$ channels and every second
$n_F=3$ channel) cannot be accommodated into a geometric structure
of supermultiplets, cf.\ Fig.\ \ref{fig:diam.nB3}.

It is also interesting to realize that, even though supersymmetry
is broken by the cutoff, the above rule is not, i.e., we don't see
any hint of continuum states in the channels $(n_f=2,4,j=odd)$ for
any finite cutoff.

\subsubsection{Scaling}

Non-localized states of the system describe D-particles \cite{SF}
penetrating the flat directions of the potential, as mentioned in the
Introduction. In a cut system all energy levels of the continuum
states fall to zero with increasing \nbm. If we label them by a
``principal quantum number'' $m$, the large-cutoff limit at fixed $m$
is trivial.  Such a phenomenon was also found in the free case when
one regularizes the system by limiting the number of quanta
\cite{TW}. In that case, it was also shown that the nontrivial and
correct continuum limit is the {\em scaling\/} limit
\eq
E(p)=\lim_{N\rightarrow \infty} E^{m(N,p)}_N,\;\;
m = \frac{p}{ \pi }\sqrt{2 N}.  \label{scaling}
\eqx
where $p$ is the continuum momentum and $N$ the cutoff. These results
were obtained analytically for a free particle in one dimension. They
also apply to the $D=2$ SU(2) SYMQM, since this is effectively a
quantum mechanics of a free particle in three (color) dimensions,
projected on the singlet and triplet channels of angular momentum
\cite{TRZ}.  The present, $D=4$, case is more complicated.
 However, we expect that,
whenever it is possible to define asymptotic states with given
momentum, as is the case for the scattering process considered
here, some version of Eq. (45) should hold. Scattering states in
the present model correspond to particles propagating freely in
the three dimensional (color) flat valleys of the potential $V$,
cf.\ Eq.\ (\ref{eq:V}).  Gauge invariance restricts color orbital
angular momentum to few channels, so we are not that far from the
$D=2$ example.  We have therefore taken Eq.\ (\ref{scaling}) as a
phenomenological rule and tested it with our data.

\begin{figure}[tbp]
\begin{center}
\epsfig{width=10cm,file=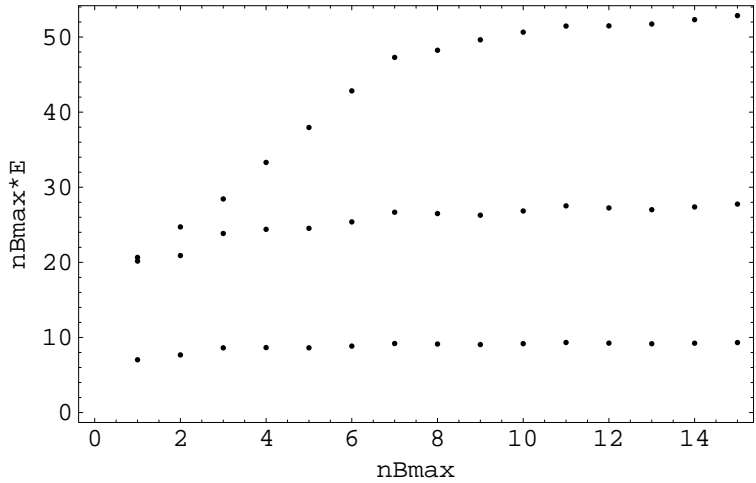}
\end{center}
\caption{ Eigenenergies of the first three levels from the continuum,
 in the $(n_F=3,j=1/2)$ channel,
 multiplied by \nbm,
as the function of the cutoff.}
\label{fig:oneoverN}
\end{figure}

The scaling limit (\ref{scaling}) implies that at fixed $m$ all
energies of non-localized states behave as ${\rm O}(1/N)$.  Figure
\ref{fig:oneoverN} tests this prediction assuming that we identify the
one dimensional cutoff $N$ with \nbm. Indeed the energies of the first
four levels seem to follow $1/n_{B_{max}}$ behavior both in $n_F=2$
and $n_F=3$ sectors.  We did not use higher levels since they are
probably influenced by the the discrete spectrum.

One can also contrast the $m$ dependence with the one dimensional
formula
\eq
E^{(m)}_N=\frac{\pi m^2}{4 N},   \label{sca}
\eqx
and with the $D=2$ case.  Table \ref{tab:Eratios} compares ratios
of our first four energy levels, for the largest value of the
cutoff, with analogous ratios of the $D=2$ system at the same
value of $N$, and with Eq.\ (\ref{sca}).

\begin{table}[tbp]
 \begin{center}
  \begin{tabular}{cccc}
\hline\hline
  $E^{(k)}/E^{(m)}$   & $D=4$ &  $D=2$  & exact   \\
               & $n_F=2,\ j=0$ &  $n_F=0$ & $\frac{k^2}{m^2}$ \\
\hline
    $E^{(2)}/E^{(1)}$  & 4.24 & 4.02 & 4.00   \\
    $E^{(3)}/E^{(1)}$  & 9.32 & 9.13 & 9.00   \\
    $E^{(3)}/E^{(2)}$  & 2.20 & 2.27 & 2.25   \\
    $E^{(4)}/E^{(1)}$  & 15.97&16.46 &16.00   \\
    $E^{(4)}/E^{(2)}$  & 3.76 & 4.09 & 4.00   \\
    $E^{(4)}/E^{(3)}$  & 1.70 & 1.80 & 1.78   \\
    \hline
         & $D=4$ & $D=2$  &                   \\
   & $n_F=3,\ j=1/2$  &  $n_F=1$    & $N=150$ \\
 \hline
    $E^{(2)}/E^{(1)}$  & 2.97 & 2.98 & 2.96   \\
    $E^{(3)}/E^{(1)}$  & 5.67 & 6.11 & 5.89   \\
    $E^{(3)}/E^{(2)}$  & 1.90 & 2.01 & 1.99   \\
    $E^{(4)}/E^{(1)}$  & 8.77 &10.18 & 9.81   \\
    $E^{(4)}/E^{(2)}$  & 2.94 & 3.42 & 3.32   \\
    $E^{(4)}/E^{(3)}$  & 1.54 & 1.69 & 1.66   \\
\hline\hline
  \end{tabular}
 \end{center}
\caption{Ratios of the energies from the continuum.
  Comparisons between the $D=2$ and $D=4$ systems. }
\label{tab:Eratios}
\end{table}

The comparison is done in two channels: the $(2,0)$ channel,
corresponding to the bosonic $n_F=0$ sector of the $D=2$ model,
and the $(3,1/2)$ channel, which is the counterpart of the
fermionic $n_F=1$ sector. \footnote{See Ref.\ \cite{CW} for the
details of the $D=2$ system} To give an idea of the cutoff
effects, we quote the $D=2$ energies for $N=18=\nBmax$ (third
column) and for $N=150$, which is easily available in this case
and coincides with $N=\infty$ within the two digits accuracy
reported (fourth column, lower half).

In the scalar case, high-cutoff results for $D=2$ are identical with
the exact ratios $k^2/m^2$.  The one-dimensional formula (\ref{sca})
does not apply to the fermionic sector. It is not surprising since
this case corresponds to color angular momentum $j=1$ and the
three dimensional Schr\"{o}dinger equation coincides with the one
dimensional one only for $j=0$.

Finally, the comparison of the ratios for $D=4$ with $D=2$ is rather
satisfactory.
Numerical values of the energy ratios for the two systems are quite similar,
over a range of an order of magnitude.
All discrepancies
are consistent with the cutoff effects. However one cannot exclude
differences $\sim 10\% $ and consequently higher \nbm\ are required
for more quantitative conclusions.

\subsubsection{Dispersion relation}

An interesting question appears whether the dispersion relation for
the scattering states has the standard parabolic form, or whether it
is modified by rather unusual behaviour of the potential.
With the help of the scaling relation (\ref{scaling}) we can now
address this issue in both bosonic and fermionic sectors.  For $n_F=2,
j=0$ the dispersion relation was first obtained by van Baal
\cite{vBN}.

\begin{figure}[tbp]
\begin{center}
\epsfig{width=14cm,file=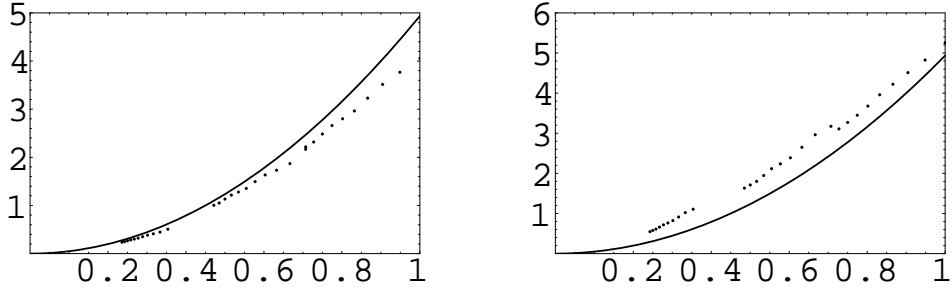}
\end{center}
\caption{ Dispersion relation for the lowest three scattering states
in the $(2,0)$ and $(3,1/2)$ channels.}
\label{fig:disp}
\end{figure}

In Fig.\ \ref{fig:disp} we have plotted the first three energy levels,
as a function of $m/\sqrt{n_{B_{max}}}$, for both bosonic ($n_F=2,
j=0$) and fermionic $(n_F=3, j=1/2)$ channels.  Points from different
$m$ and \nbm\ follow roughly a common curve which again confirms
approximately the scaling relation (\ref{scaling}). Moreover, when the
proper normalization of the momentum, required in (\ref{scaling}), is
taken into account, one obtains a reasonable agreement with the
standard $p^2/2$ kinetic energy of one degree of freedom (solid
lines).

Many effects prevent us from reaching better agreement at the moment.    For
example, the repulsion of the lowest discrete state at $E=4.12$ is
clearly seen in the (2,0) channel, while it is not as efficient in
(3,1/2), where the lowest state is higher ($E=5.19$). For the present
values of \nbm, only the three lowest states can be used, hence one
expects non-leading corrections in $m$.  The identification of $N$
with \nbm\ should be more carefully examined, etc.  However,
keeping in mind all these limitations, the overall picture seems
reasonably satisfactory and we are looking forward for better data to
make more extensive study of these points.

\subsection{Witten index}

\begin{figure}[tbp]
\begin{center}
\psfig{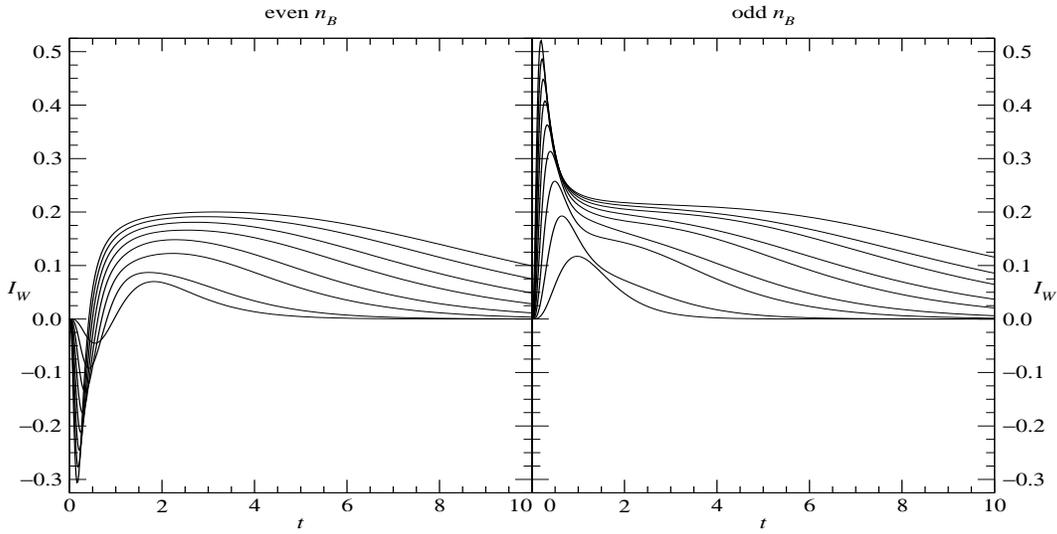}\qquad
\end{center}
\caption{Regularized Witten index $I_W(T)$ vs.\ $T$ for even (left)
and odd (right) $\nBmax\le17$ .}
\label{fig:Iw}
\end{figure}

With complete diagonalization of the Hamiltonian achieved in all
sectors we can now calculate the regularized Witten index directly
from the definition
\[ I_W(T)=\sum_{i} (-1)^{n_F(i)} e^{-T E(i)}. \]
The results, shown in Fig.~\ref{fig:Iw}, nicely confirm and strengthen
early expectations based on much smaller \nbm.

As already mentioned, the number of bosonic and fermionic states is
the same for any value of the cutoff, be it even or odd. Therefore the
index vanishes at $T=0$ with this regularization. The sharp
structure around $T=0$ clearly moves toward the origin with
increasing cutoff indicating singularity at infinite $\nBmax$.
Such a discontinuity is expected on general grounds and finds a
reasonable support here.

Of course at high $T$ our ``cut'' index is bound to vanish
exponentially.  However there exist a range of intermediate times
where definite flattening occurs.  This signals effective
cancellations among supersymmetric partners hence a gradual,
global restoration of SUSY.  Moreover the plateau seems to
converge to $\fourth$ --- a known result obtained also from the
non-abelian integrals for the SU(2) gauge group \cite{SM} -
\cite{ST}.

\begin{figure}[tbp]
\begin{center}
\psfig{file=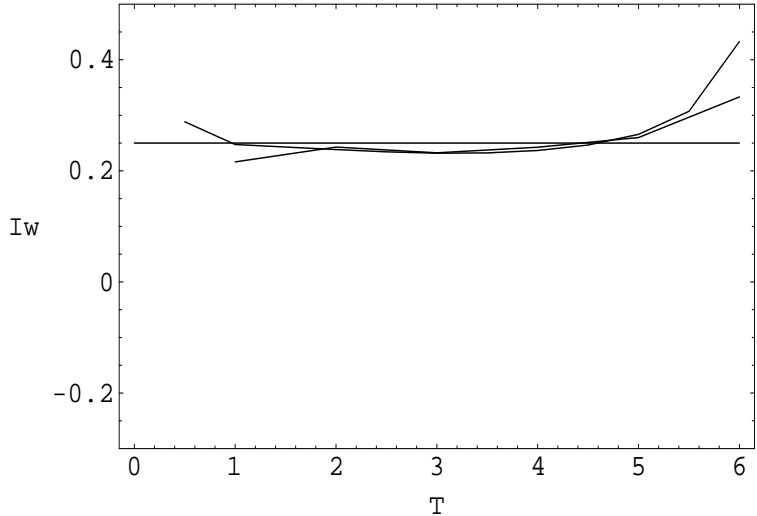,width=10cm, height=7cm}\qquad
\end{center}
\caption{Infinite \nbm limit of the index deduced from the
Pad\`{e} approximants.} \label{fig:pade}
\end{figure}

In order to study the large cutoff limit more quantitatively we have performed
a number of extrapolations assuming various
asymptotic behaviors of the regularized index in \nbm.
For example, Fig. \ref{fig:pade} shows the asymptotic
value extracted with the aid of the diagonal Pad\`{e} approximant   $P_{[4,4]}(\nBmax^2)$
at various $T$. Two lines correspond to even and odd \nbm cases, which were
independently analyzed .
Both extrapolations are stable and consistent in the range $1<T<5$.
This result strongly suggests that the infinite \nbm limit of the regularized index
is time independent.
Moreover, the limiting value is nicely consistent with the above $1/4$
(also shown in the Figure).
Extrapolations
with power series in different variables lead to similar conclusions.
We expect to accumulate new data with yet higher cutoffs. This would allow to
extend stable extrapolations to larger range of $T$ and possibly distinguish between
various asymptotic forms tested so far.

\section{Summary and outlook}

The next step in studying a family of supersymmetric Yang-Mills
quantum mechanics has been completed. The above models appear in
many areas of theoretical physics, beginning with the soluble
$D=2$ systems, through the small volume, lattice studies of the
$D=4$ QCD, and finally ending on the $D=10$ models of the
M-theory. We are now somewhere in the middle of this list.

The new approach presented here leads to the precise study of the rich
structure of the $D=4$ system, which already has some features of the
$D=10$ model.
With rotational invariance taken fully into account,
the Hilbert space splits into channels of conserved angular momentum and fermionic
number. This eliminates the brute force diagonalization of large matrices. Second improvement is
brought by the generalization of the recursive scheme of computing matrix elements
while gradually increasing the harmonic oscillator basis \cite{CW}. Present results fully confirm and
extend findings of the first paper where the whole program was originated \cite{JWNP}.

     The system has both discrete and continuous spectrum which coexist at the same energies. This rather
unusual feature was expected for a long time as a consequence of the supersymmetric
interactions with flat directions. Now however, more precise statements can be made. While the discrete,
localized states exist in all $(n_F,j)$ channels, the non-localized ones appear only
in the central (with respect to the particle-whole reflection) supermultiplets and only
for even angular momenta.

The numerical part of the method requires limiting the Hilbert space. We take as a cutoff the
maximal number of quanta of all bosonic harmonic oscillators,
$\nBmax$. The present approach allows to reach
such a large cutoffs that the lower part of the discrete spectrum has practically converged
to its continuum (i.e., the infinite cutoff) limit.

On the other hand the eigenenergies from the continuous spectrum
literally never converge to their continuum values. Instead, they
all fall to zero with increasing cutoff. In fact, this is precisely
the property allowing a clear distinction of the two spectra, cf.\ Figs
\ref{fig:vanBaal} and \ref{fig:Ecombined}. The physical energies
of the non-localized states are coded in the rate of fall of the above
levels with $\nBmax$. The particular scaling which governs this
behavior was discovered some time ago \cite{TW} and is well
confirmed with present data. It is an important tool in extracting
any observable related to the non-localized states. In particular
it allowed us to establish
 the dispersion relation
for the scattering states in the $(n_F=2,j=0)$ and $(n_F=3,j=\half)$ channels.

      Supersymmetry is broken by the cutoff. Again however, with currently available values of
$\nBmax$, we observe clear restoration of SUSY which manifests itself in many ways in the discrete
spectrum. First, the energy levels from different channels,
related by supersymmetry, coincide to high
accuracy, cf.\ Fig.\ \ref{allFs} .
 Second, our approach allows to form and analyze the supersymmetric images
of arbitrary eigenstates. This led to the construction of the rotationally
 invariant supersymmetry fractions which provided a simple identification
 of SUSY partners. A number of lower supermultiplets was identified for a range
 of angular momenta, see Tables\ \ref{tab:spectrum} and \ref{tab:spectrum2}.
Interestingly some of the supermultiplets are degenerate, see
Fig.\ \ref{fig3D}. We do not know a symmetry (if any) responsible
for this additional degeneracy. The mixing angles are stable with
respect to changing the cutoff.
Their actual values, however, may be an artefact of our
regularization.

        A third method to see restoration of supersymmetry is provided by the Witten index.
It is clearly flattening as a function of euclidean time when we move towards bigger cutoffs which are now
available. This shows that the cancellations between supersymmetric partners becomes more and more
efficient, also globally. At infinite cutoff contribution from localized states would be exactly zero.
Supersymmetric vacuum and other non-localized states should give the final non-integer value 1/4 for
the gauge group considered here. We see
now much stronger evidence for this behavior than in the first attempts.

      In the continuous sector of the theory the situation is more difficult and challenging.
Although the scaling expected from the one dimensional free case
has been confirmed, it should be studied now more extensively,
also for higher angular momenta. Identification of the
supermultiplets is more delicate and remains to be done. Similarly
revealing a signature of the SUSY vacuum requires further study
and yet higher cutoffs. On the other hand current precision allows
to address more advanced problems like the scattering
\cite{PW,BERS}. We are looking forward to work out some of these
questions.

Progress towards higher $\nBmax$ for the $D=4$ system is limited by
computer time.  The recursive algorithm is presently implemented in
Mathematica.  We re-implemented some sections of the algorithm in C++,
obtaining a 100-fold increase in speed; we plan to complete the C++
implementation and to improve the present computation.

Altogether the present approach works rather well.
As such it provides one route of attacking
higher dimensions.  The gain from exploiting fully SO($D{-}1$)
invariance and
 restricting ourself to
a particular representation of SO($D{-}1$) should overcome the huge sizes of bases in higher dimensions.
Generalization to $D=10$ requires in particular construction of the Clebsch-Gordan
coefficients for the SO(9) group which is a reasonably tedious but a well defined exercise.
Some work in this direction has already begun.

\vspace*{0.5cm}

\noindent {\em Acknowledgments.} We would like to thank P. van
Baal for discussions and for providing the data for Fig.2.  This
work is supported by the Polish Committee for Scientific Research
under the grants no.\ 2~P03B~096~22 (2002-2004) and 1~ P03B~024~27
(2004-2007), and by INFN under IS~PI12.

\appendix

\section{Useful identities involving Clebsch-Gordan coefficients,
$\bm{3j}$ symbols, and $\bm{6j}$ symbols}
\label{app:6j}

With the usual phase conventions, the Clebsch-Gordan coefficients
$C^{j_1\,j_2\,j}_{m_1m_2m}$ are real and
\begin{equation}
C^{j\,j\,0}_{m_1m_20} =
{(-1)^{j-m_1}\over\sqrt{2j+1}}\,\delta_{m_1+m_2,0};
\label{eq:Cll0}
\end{equation}
the completeness formulae read
\begin{equation}
\sum_{j,m} C^{j_1\,j_2\,j}_{m'_1m'_2m}  C^{j_1\,j_2\,j}_{m_1m_2m}
   = \delta_{m'_1m_1}\delta_{m'_2m_2}, \qquad
\sum_{m_1,m_2} C^{j_1\,j_2\,j'}_{m_1m_2m'}  C^{j_1\,j_2\,j}_{m_1m_2m}
   = \delta_{m'm}\delta_{j'j}.
\label{eq:CGcompl}
\end{equation}

The Clebsch-Gordan coefficients can be written in terms of the Wigner
$3j$ symbols as
\begin{equation}
C^{j_1\,j_2\,j}_{m_1m_2m} =
(-1)^{j_1-j_2+m} \, \sqrt{2j+1} \threej{j_1&j_2&j \\ m_1&m_2&-m};
\label{eq:3j}
\end{equation}
the $3j$ symbols enjoy the symmetry properties
\begin{eqnarray}
\threej{j_2&j_1&j_3 \\ m_2&m_1&m_3} &=&
(-1)^{j_1+j_2+j_3}\threej{j_1&j_2&j_3 \\ m_1&m_2&m_3},
\label{eq:3j-exch} \\
\threej{j_2&j_1&j_3 \\ -m_1&-m_2&-m_3} &=&
(-1)^{j_1+j_2+j_3}\threej{j_1&j_2&j_3 \\ m_1&m_2&m_3};
\label{eq:3j-inv}
\end{eqnarray}
we also need the formula \cite{Landau-III}
\begin{eqnarray}
&&  \sum_{m_4,m_5,m_6} (-1)^{j_4+j_5+j_6-m_4-m_5-m_6}
    \threej{j_1&j_5&j_6 \\ m_1&-m_5&m_6}
    \threej{j_4&j_2&j_6 \\ m_4&m_2&-m_6}
    \threej{j_4&j_5&j_3 \\ -m_4&m_5&m_3} \nonumber \\
&& \qquad =
    \threej{j_1&j_2&j_3 \\ m_1&m_2&m_3}
    \sixj{j_1&j_2&j_3 \\ j_4&j_5&j_6} ,
\label{eq:6j}
\end{eqnarray}
where the term in braces is the Racah $6j$ symbol.
Eqs.\ (\ref{eq:3j}), (\ref{eq:3j-exch}), and (\ref{eq:3j-inv})
imply the ``exchange'' formulae
\begin{equation}
C^{j_2\,j_1\,j}_{m_2m_1m} =
   (-1)^{j_1+j_2-j} \, C^{j_1\,j_2\,j}_{m_1m_2m}, \qquad
C^{j_1\,j\,j_2}_{m_1mm_2} = (-1)^{j-j_2-m_1} \,
   \sqrt{2j_2+1\over2j+1} \, C^{\;j_1\,j_2\,j}_{-m_1m_2m},
\label{eq:CGexch}
\end{equation}
and the ``inversion'' formula
\begin{equation}
C^{j_2\ j_1\ j}_{-m_2\,{-}m_1\,{-}m} =
   (-1)^{j_1+j_2-j} \, C^{j_1\,j_2\,j}_{m_1m_2m};
\label{eq:CGinv}
\end{equation}
Eq.\ (\ref{eq:6j}) implies
\begin{eqnarray}
&& \sum_{m_4,m_5,m_6} C^{j_4\,j_5\,j_1}_{m_4m_5m_1} \,
   C^{j_4\,j_6\,j_3}_{m_4m_6m_3} \, C^{j_5\,j_2\,j_6}_{m_5m_2m_6} \nonumber \\
&& \qquad = (-1)^{j_2+j_3+j_4+j_5} \, \sqrt{(2j_1+1)(2j_6+1)}
   \sixj{j_1&j_2&j_3 \\ j_6&j_4&j_5}
   C^{j_1\,j_2\,j_3}_{m_1m_2m_3}.
\label{eq:CG3}
\end{eqnarray}

\section{Computation of matrix elements of products of operators}
\label{app:W-E}

We wish to exploit rotation invariance to reduce the number of
matrix elements which must be computed and stored; our main tool is
the Wigner-Eckhart theorem:
\begin{equation}
\langle j',m';i'|{\cal O}^{(1)}_{j_1,m_1}|j,m;i\rangle =
{1\over\sqrt{2j'+1}}\,C^{j_1\,j\,j'}_{m_1mm'}
{\langle j';i'\Vert {\cal O}^{(1)}_{j_1}\Vert j;i\rangle},
\label{eq:J-ME}
\end{equation}
where $\langle j';i'\Vert {\cal O}^{(1)}_{j_1}\Vert j;i\rangle$
denotes a {\em reduced matrix element}, independent on $m'$, $m_1$,
and $m$.

A first formula regarding reduced matrix element can be easily
obtained applying Eqs.\ (\ref{eq:CGexch}) and (\ref{eq:CGinv}):
\begin{equation}
{\langle j';i'\Vert  {\tilde{\cal O}}_{j_1}\Vert j;i\rangle}
= (-1)^{j_1+j-j'} \,
{\langle j;i\Vert  {\cal O}^\dagger_{j_1}\Vert j';i'\rangle}, \qquad
{\tilde{\cal O}}_{j,m} = (-1)^{j+m} \, {\cal O}_{j,-m}.
\label{eq:redadj}
\end{equation}

A second formula deals with the product of two operators ${\cal
O}^{(1)}_{j_1,m_1}$ and ${\cal O}^{(2)}_{j_2,m_2}$ with given bosonic
and fermionic number: by decomposing the of the product in components
with definite $\bm{J}^2$ and then applying Eq.\ (\ref{eq:CG3}), we
obtain
\begin{eqnarray}
&& {\langle j';i'\Vert \bigl({\cal O}^{(1)}_{j_1},
   {\cal O}^{(2)}_{j_2}\bigr)_{j_3}\Vert j;i\rangle} \nonumber \\
&&\quad =
\sum_{j'',i''} (-1)^{j+j'+j_1+j_2} \, \sqrt{2j_3+1}
   \sixj{j_3&j&j' \\ j''&j_1&j_2}
{\langle j';i'\Vert {\cal O}^{(1)}_{j_1}\Vert j'';i''\rangle} \;
{\langle j'';i''\Vert {\cal O}^{(2)}_{j_2}\Vert j;i\rangle}.
\label{eq:redprod}
\end{eqnarray}

Applying Eqs.\ (\ref{eq:redadj}) and (\ref{eq:redprod}), we never need
to deal explicitly with $m$'s and Clebsch-Gordan coefficients, and the
computations are much more efficient.

(Anti)commutators of operators are dealt with in a very similar way.

\section{Implementation of the algorithm}
\label{app:algo}

We implemented our algorithms in the symbolic manipulation program
Mathematica.

We first compute tables of all needed (anti)commutators and
decompositions of products of two trilinear operators into products of
three bilinear operators: we define an explicit and univocal
representation of a generic operator in terms of $a$, $a^\dagger$,
$f$, $f^\dagger$, in a ``canonical'' order; using this representation,
we compute explicitly the desired operators and decompose them in the
appropriate basis of gauge-invariant operators.  The computation so
far is exact, and the coefficients are square roots of rational
numbers.  Many checks are performed: besides verifying the rotational
properties, we check explicitly Eqs.\ (\ref{eq:HK}), (\ref{eq:HP}),
(\ref{eq:HF}), (\ref{eq:QH}), and (\ref{eq:Q-def}).  This step
requires moderate computer resources.  Once the tables are computed,
the explicit representation of the gauge-invariant operators is no
longer needed.

A separate program reads in the tables and implements the
orthonormalization and recursive computation of scalar products and
matrix elements described in Sects.\ \ref{sec:ortho} and
\ref{sec:recursive}.  The formulae given in the two sections, together
with Hermiticity and decompositions of products of two trilinear
operators, are more than sufficient to reduce any matrix element to
matrix elements involving a lower number of elementary creation and
annihilation operators; in many instances, more than one reduction is
available, and the choice can affect performance very strongly.

For performance reasons, it is crucial to ``remember'' the values of
all matrix elements already computed, and to save them periodically
into a file to be able to restart the computation.  Again for
performance reasons, we choose to represent matrix elements as
double-precision floating point numbers rather than as exact
algebraic numbers.

\section{Sizes of bases}
\label{app:size}

The algorithm described in Sec.\ \ref{sec:newalgo} generates bases in each
channel $(n_F,j)$, recursively in $n_B$, by applying all operators
listed in Secs \ \ref{sec:bosonic}, \ref{sec:fermionic}. Then the Gram-Schmidt
orthonormalization
selects maximal set of linearly independent states. Their numbers
are quoted in Tables \ref{tableF0}-\ref{tableF3}.

Equivalently, dimensions of the above-mentioned subspaces can be
derived by classifying all independent tensor structures
contributing to each channel at given $n_B$. This provides an
additional check of our program and prepares the ground for
subsequent generalization to higher dimensions and higher gauge
groups. Here are few examples for various $n_F$.\newline
 {{\bf D.1} All $n_F=0$ states.}\newline
 For even $n_B$, every gauge-invariant
state can be obtained by applying to the vacuum a combination of
the gauge-invariant creation operators
 \eq
 A^{ik}=a_b^{\dagger i}a_b^{\dagger k}. \label{eq:aa}
 \eqx
Since there are 6 independent $A^\dagger_{ik}$, and states created by
different products of $A^\dagger_{ik}$, apart from permutations, are
linearly independent, the total size of the basis with even number of bosons is
\[ {\cal N}(n_F{=}0,n_B{=}2n) = {n+5\choose n}. \]

 \noindent {{\bf D.2} $n_F=0, j=0,2.$}\newline
 All gauge invariant and spherically symmetric states can be obtained
 by combining the traces of the powers of the basic gauge invariant bilinear creator
(\ref{eq:aa}).
 Since A is a three by three matrix, its Cayley-Hamilton equation is third order, hence only
 traces of first three powers of $A$ are independent. It follows that the number of independent states
with $n_B$ quanta equals to the number of monomials of the $n_B/2$
order which can be made from $\Tr A$, $\Tr(A^2)$ and $\Tr(A^3)$.
Therefore it is given by the number of partitions $P(n_B/2|1,2,3)$
\eq {\cal N}(0,0,n_B)=P(n_B/2|1,2,3),\; n_B - even,
\label{eq:NF0L0Be} \eqx of $n_B/2$ into elements smaller than $4$.

States with odd $n_B$ can be generated by acting with the only odd
creator ${\bar A}^\dagger$, Eq.\ (\ref{eq:AAAdag}), on the even
basis. Therefore

 \eq
{\cal N}(j=0,n_F=0,n_B)={\cal N}(0,0,n_B-3),\;\; n_B - odd.
\label{eq:nBodd} \eqx

This explains the even-odd regularities in the first column of
Table \ref{tableF0}. Since ${\bar A}^\dagger$ is a scalar
Eq.(\ref{eq:nBodd}) holds for arbitrary angular momentum $j$ and
consequently also for global number of states, cf.\ Table
\ref{tab:size-vs-nB}.

 \begin{landscape}
  \begin{table}[tbp]
    \begin{tabular}{ccccccccccccccccccccrr} \hline\hline
    $J$ & $ 0 $ & $ 1 $ & $ 2 $ & $ 3 $  & $ 4 $ & $ 5 $ & $ 6 $ & $ 7 $  & $ 8 $ & $ 9 $ & $ 10 $ & $ 11 $  & $ 12 $ & $ 13 $ & $ 14 $ & $ 15 $  & $ 16 $ & $ 17 $  & $ 18 $ & $ N_s $ & $ \Sigma $  \\
    $n_B$ & $   $ & $   $ & $   $ & $   $  & $   $ & $   $ & $   $ & $   $  & $   $ & $   $ & $    $ & $    $  & $    $ & $    $ & $    $ & $    $  & $    $ & $    $  & $    $ & $     $ & $        $  \\
    \hline
    $0$  & $ 1 $ & $   $ & $   $ & $   $  & $   $ & $   $ & $   $ & $   $  & $   $ & $   $ & $    $ & $    $  & $    $ & $    $ & $    $ & $    $  & $    $ & $    $  & $    $ & $  1  $ & $    1   $  \\
    $1$  & $ 0 $ & $ 0 $ & $   $ & $   $  & $   $ & $   $ & $   $ & $   $  & $   $ & $   $ & $    $ & $    $  & $    $ & $    $ & $    $ & $    $  & $    $ & $    $  & $    $ & $  -  $ & $    1   $  \\
    $2$  & $ 1 $ & $ 0 $ & $ 1 $ & $   $  & $   $ & $   $ & $   $ & $   $  & $   $ & $   $ & $    $ & $    $  & $    $ & $    $ & $    $ & $    $  & $    $ & $    $  & $    $ & $  6  $ & $    7   $  \\
    $3$  & $ 1 $ & $ 0 $ & $ 0 $ & $ 0 $  & $   $ & $   $ & $   $ & $   $  & $   $ & $   $ & $    $ & $    $  & $    $ & $    $ & $    $ & $    $  & $    $ & $    $  & $    $ & $  1  $ & $    8   $  \\
    $4$  & $ 2 $ & $ 0 $ & $ 2 $ & $ 0 $  & $ 1 $ & $   $ & $   $ & $   $  & $   $ & $   $ & $    $ & $    $  & $    $ & $    $ & $    $ & $    $  & $    $ & $    $  & $    $ & $ 21  $ & $   29   $  \\
    $5$  & $ 1 $ & $ 0 $ & $ 1 $ & $ 0 $  & $ 0 $ & $ 0 $ & $   $ & $   $  & $   $ & $   $ & $    $ & $    $  & $    $ & $    $ & $    $ & $    $  & $    $ & $    $  & $    $ & $  6  $ & $   35   $  \\
    $6$  & $ 3 $ & $ 0 $ & $ 3 $ & $ 1 $  & $ 2 $ & $ 0 $ & $ 1 $ & $   $  & $   $ & $   $ & $    $ & $    $  & $    $ & $    $ & $    $ & $    $  & $    $ & $    $  & $    $ & $ 56  $ & $   91   $  \\
    $7$  & $ 2 $ & $ 0 $ & $ 2 $ & $ 0 $  & $ 1 $ & $ 0 $ & $ 0 $ & $ 0 $  & $   $ & $   $ & $    $ & $    $  & $    $ & $    $ & $    $ & $    $  & $    $ & $    $  & $    $ & $ 21  $ & $  112   $  \\
    $8$  & $ 4 $ & $ 0 $ & $ 5 $ & $ 1 $  & $ 4 $ & $ 1 $ & $ 2 $ & $ 0 $  & $ 1 $ & $   $ & $    $ & $    $  & $    $ & $    $ & $    $ & $    $  & $    $ & $    $  & $    $ & $ 126 $ & $  238   $  \\
    $9$  & $ 3 $ & $ 0 $ & $ 3 $ & $ 1 $  & $ 2 $ & $ 0 $ & $ 1 $ & $ 0 $  & $ 0 $ & $ 0 $ & $    $ & $    $  & $    $ & $    $ & $    $ & $    $  & $    $ & $    $  & $    $ & $  56 $ & $  294   $  \\
    $10$ & $ 5 $ & $ 0 $ & $ 7 $ & $ 2 $  & $ 6 $ & $ 2 $ & $ 4 $ & $ 1 $  & $ 2 $ & $ 0 $ & $ 1  $ & $    $  & $    $ & $    $ & $    $ & $    $  & $    $ & $    $  & $    $ & $ 252 $ & $  546   $  \\
    $11$ & $ 4 $ & $ 0 $ & $ 5 $ & $ 1 $  & $ 4 $ & $ 1 $ & $ 2 $ & $ 0 $  & $ 1 $ & $ 0 $ & $ 0  $ & $ 0  $  & $    $ & $    $ & $    $ & $    $  & $    $ & $    $  & $    $ & $ 126 $ & $  672   $  \\
    $12$ & $ 7 $ & $ 0 $ & $ 9 $ & $ 3 $  & $ 9 $ & $ 3 $ & $ 7 $ & $ 2 $  & $ 4 $ & $ 1 $ & $ 2  $ & $ 0  $  & $  1 $ & $    $ & $    $ & $    $  & $    $ & $    $  & $    $ & $ 462 $ & $ 1134   $  \\
    $13$ & $ 5 $ & $ 0 $ & $ 7 $ & $ 2 $  & $ 6 $ & $ 2 $ & $ 4 $ & $ 1 $  & $ 2 $ & $ 0 $ & $ 1  $ & $ 0  $  & $  0 $ & $  0 $ & $    $ & $    $  & $    $ & $    $  & $    $ & $ 252 $ & $ 1386   $  \\
    $14$ & $ 8 $ & $ 0 $ & $12 $ & $ 4 $  & $12 $ & $ 5 $ & $10 $ & $ 4 $  & $ 7 $ & $ 2 $ & $ 4  $ & $ 1  $  & $  2 $ & $  0 $ & $ 1  $ & $    $  & $    $ & $    $  & $    $ & $ 792 $ & $ 2178   $  \\
    $15$ & $ 7 $ & $ 0 $ & $ 9 $ & $ 3 $  & $ 9 $ & $ 3 $ & $ 7 $ & $ 2 $  & $ 4 $ & $ 1 $ & $ 2  $ & $ 0  $  & $  1 $ & $  0 $ & $ 0  $ & $ 0  $  & $    $ & $    $  & $    $ & $ 462 $ & $ 2640   $  \\
    $16$ & $10 $ & $ 0 $ & $15 $ & $ 5 $  & $16 $ & $ 7 $ & $14 $ & $ 6 $  & $11 $ & $ 4 $ & $ 7  $ & $ 2  $  & $  4 $ & $  1 $ & $ 2  $ & $ 0  $  & $ 1  $ & $    $  & $    $ & $1287 $ & $ 3927   $  \\
    $17$ & $ 8 $ & $ 0 $ & $12 $ & $ 4 $  & $12 $ & $ 5 $ & $10 $ & $ 4 $  & $ 7 $ & $ 2 $ & $ 4  $ & $ 1  $  & $  2 $ & $  0 $ & $ 1  $ & $ 0  $  & $ 0  $ & $ 0  $  & $    $ & $ 792 $ & $ 4719   $  \\
    $18$ & $12 $ & $ 0 $ & $18 $ & $ 7 $  & $20 $ & $ 9 $ & $19 $ & $ 9 $  & $15 $ & $ 7 $ & $11  $ & $ 4  $  & $  7 $ & $  2 $ & $ 4  $ & $ 1  $  & $ 2  $ & $ 0  $  & $ 1  $ & $2002 $ & $ 6721   $  \\
 \hline\hline
    \end{tabular}
    \caption{Number of SO(3) multiplets with $n_F=0$ and fixed $j$ and $n_B$. $N_s$ is the number of basis vectors
   with given number of bosonic quanta, $n_B$, while $\Sigma$ gives the cumulative size up to $n_B$.}
 \label{tableF0}
 \end{table}
\end{landscape}

The $j=2$ states can be generated from the empty states by replacing one of the two traces
$\Tr(A)$  or $\Tr(A^2)$ by the symmetric traceless tensor formed from $A$ or $A^2$. Note that
such a tensor formed from $A^3$ is already dependent on the lower powers of $A$. This is again the
consequence of the Cayley-Hamilton equation: only the {\em trace} of the $A^3$ is independent since
it is equivalent to one of the coefficients of the C-H equation (namely to the determinant). One can
therefore count the j=2 states as follows: for each even $n_B$ take all partitions contributing to
Eq.\ (\ref{eq:NF0L0Be}), replace in each partition one element, e.g., 2 by its indexed counterpart $2^{ik}$.
This produces a monomial with $\Tr(A^2) \rightarrow (A^2)^{ik}-A^2\delta^{ik}$ which generates one
$j=2$ state. Repeat this procedure for all different elements in a partition omitting value 3.
Total number of states equals to the number of such indexed monomials. This procedure
 indeed reproduces sizes listed in the third column of Table\ref{tableF0}. Yet simpler counting can be
formulated recursively: states with $n_B$ quanta can be obtained by acting with $A^{ik}$ on the $j=0,n_b-2$
basis and independently by acting with $(A^2)^{ik}$ on the $j=0,n_B-4$ basis. This gives the
recursion relation
\eq
{\cal N}(0,2,n_B)={\cal N}(0,0,n_B-2)+{\cal N}(0,0,n_B-4),
\eqx
which explains the $j=2$ column of Table\ref{tableF0}.

\noindent {\bf D.3} $n_F=1, j=1/2,3/2.$\newline
The lowest gauge
invariant state in this sector must contain one boson and is
created by $f_b^{\dagger\sigma}a_b^{\dagger i}\equiv (fa)^{\sigma
i}$ from the empty state. States with $j=1/2$ and $j=2/3$ are
generated  by suitable projections $(fa)_{1/2}$ and $(fa)_{3/2}$,
where $(...)_j$ means summing over $\sigma$ and $i$ indices with
appropriate Clebsch-Gordan coefficients. For odd $n_B$ one then
combines powers of traces of $A$,$A^2$ and $A^3$ from previous
case with three independent\footnote{Again Cayley-Hamilton
 equation for A limits a number of independent creators with $n_F=1$.} fermionic creators $(fa)_{1/2},\;
 (fa.A)_{1/2},\;(fa.A^2)_{1/2}$ to get all states
with $j=1/2$ \footnote{A "." denotes contraction of an adjoint
SO(3) indices.}. One can generate all states of the $n_B$ basis
recursively by acting with $(fa)_{1/2}$ on the $n_F=0,j=0, n_B-1$
basis; with $(fa.A)_{1/2}$ on the $n_F=0,j=0, n_B-3$ basis;
 and $(fa.A^2)_{1/2}$ on the
$n_F=0,j=0, n_B-5$ basis. This implies the relation \eq {\cal
N}(1,1/2,n_B)={\cal N}(0,0,n_B-1)+{\cal N}(0,0,n_B-3)+{\cal
N}(0,0,n_B-5), \eqx which gives the first column of Table
\ref{tableF1} in terms of Table \ref{tableF0}.

  \begin{landscape}
   \begin{table}[tbp]
    \begin{tabular}{ccccccccccccccccccccrr} \hline\hline
    $J$ & $ {1\over 2} $ & ${3\over 2} $ & ${5\over 2} $ & ${7\over 2}$ & ${9\over 2}$ & ${11\over 2}$ & $ {13\over 2}$ & $ {15\over 2}$  & $ {17\over 2}$ & $ {19\over 2}$ & $ {21\over 2}$ & ${23\over 2}$  & ${25\over 2}$ & ${27\over 2}$ & ${29\over 2}$ & $ {31\over 2}$  & $ {33\over 2}$ & $ {35\over 2}$  & $ {37\over 2}$ & $ N_s $ & $ \Sigma $  \\
    $n_B$ & $   $ & $   $ & $   $ & $   $  & $   $ & $   $ & $   $ & $   $  & $   $ & $   $ & $    $ & $    $  & $    $ & $    $ & $    $ & $    $  & $    $ & $    $  & $    $ & $     $ & $        $  \\
    \hline
    $0$  & $ 0 $ & $   $ & $   $ & $   $  & $   $ & $   $ & $   $ & $   $  & $   $ & $   $ & $    $ & $    $  & $    $ & $    $ & $    $ & $    $  & $    $ & $    $  & $    $ & $  0  $ & $    0   $  \\
    $1$  & $ 1 $ & $ 1 $ & $   $ & $   $  & $   $ & $   $ & $   $ & $   $  & $   $ & $   $ & $    $ & $    $  & $    $ & $    $ & $    $ & $    $  & $    $ & $    $  & $    $ & $  6  $ & $    6   $  \\
    $2$  & $ 1 $ & $ 1 $ & $ 0 $ & $   $  & $   $ & $   $ & $   $ & $   $  & $   $ & $   $ & $    $ & $    $  & $    $ & $    $ & $    $ & $    $  & $    $ & $    $  & $    $ & $  6  $ & $   12   $  \\
    $3$  & $ 2 $ & $ 3 $ & $ 2 $ & $ 1 $  & $   $ & $   $ & $   $ & $   $  & $   $ & $   $ & $    $ & $    $  & $    $ & $    $ & $    $ & $    $  & $    $ & $    $  & $    $ & $ 36  $ & $   48   $  \\
    $4$  & $ 2 $ & $ 3 $ & $ 2 $ & $ 1 $  & $ 0 $ & $   $ & $   $ & $   $  & $   $ & $   $ & $    $ & $    $  & $    $ & $    $ & $    $ & $    $  & $    $ & $    $  & $    $ & $ 36  $ & $   84   $  \\
    $5$  & $ 4 $ & $ 6 $ & $ 5 $ & $ 4 $  & $ 2 $ & $ 1 $ & $   $ & $   $  & $   $ & $   $ & $    $ & $    $  & $    $ & $    $ & $    $ & $    $  & $    $ & $    $  & $    $ & $126  $ & $  210   $  \\
    $6$  & $ 4 $ & $ 6 $ & $ 5 $ & $ 4 $  & $ 2 $ & $ 1 $ & $ 0 $ & $   $  & $   $ & $   $ & $    $ & $    $  & $    $ & $    $ & $    $ & $    $  & $    $ & $    $  & $    $ & $126  $ & $  336   $  \\
    $7$  & $ 6 $ & $ 10$ & $ 10$ & $ 9 $  & $ 6 $ & $ 4 $ & $ 2 $ & $ 1 $  & $   $ & $   $ & $    $ & $    $  & $    $ & $    $ & $    $ & $    $  & $    $ & $    $  & $    $ & $336  $ & $  672   $  \\
    $8$  & $ 6 $ & $ 10$ & $ 10$ & $ 9 $  & $ 6 $ & $ 4 $ & $ 2 $ & $ 1 $  & $ 0 $ & $   $ & $    $ & $    $  & $    $ & $    $ & $    $ & $    $  & $    $ & $    $  & $    $ & $336  $ & $ 1008   $  \\
    $9$  & $ 9 $ & $ 15$ & $ 16$ & $ 16$  & $ 13$ & $ 10$ & $ 6 $ & $ 4 $  & $ 2 $ & $ 1 $ & $    $ & $    $  & $    $ & $    $ & $    $ & $    $  & $    $ & $    $  & $    $ & $756  $ & $ 1764   $  \\
    $10$ & $ 9 $ & $ 15$ & $ 16$ & $ 16$  & $ 13$ & $ 10$ & $ 6 $ & $ 4 $  & $ 2 $ & $ 1 $ & $ 0  $ & $    $  & $    $ & $    $ & $    $ & $    $  & $    $ & $    $  & $    $ & $756  $ & $ 2520   $  \\
    $11$ & $ 12$ & $ 21$ & $ 24$ & $ 25$  & $ 22$ & $ 19$ & $ 14$ & $ 10$  & $ 6 $ & $ 4 $ & $ 2  $ & $ 1  $  & $    $ & $    $ & $    $ & $    $  & $    $ & $    $  & $    $ & $1512 $ & $ 4032   $  \\
    $12$ & $ 12$ & $ 21$ & $ 24$ & $ 25$  & $ 22$ & $ 19$ & $ 14$ & $ 10$  & $ 6 $ & $ 4 $ & $ 2  $ & $ 1  $  & $  0 $ & $    $ & $    $ & $    $  & $    $ & $    $  & $    $ & $1512 $ & $ 5544   $  \\
    $13$ & $ 16$ & $ 28$ & $ 33$ & $ 36$  & $ 34$ & $ 31$ & $ 25$ & $ 20$  & $ 14$ & $ 10$ & $ 6  $ & $ 4  $  & $  2 $ & $  1 $ & $    $ & $    $  & $    $ & $    $  & $    $ & $2772 $ & $ 8316   $  \\
    $14$ & $ 16$ & $ 28$ & $ 33$ & $ 36$  & $ 34$ & $ 31$ & $ 25$ & $ 20$  & $ 14$ & $ 10$ & $ 6  $ & $ 4  $  & $  2 $ & $  1 $ & $ 0  $ & $    $  & $    $ & $    $  & $    $ & $2772 $ & $11088   $  \\
    $15$ & $ 20$ & $ 36$ & $ 44$ & $ 49$  & $ 48$ & $ 46$ & $ 40$ & $ 34$  & $ 26$ & $ 20$ & $ 14 $ & $ 10 $  & $  6 $ & $  4 $ & $ 2  $ & $ 1  $  & $    $ & $    $  & $    $ & $4752 $ & $15840   $  \\
    $16$ & $ 20$ & $ 36$ & $ 44$ & $ 49$  & $ 48$ & $ 46$ & $ 40$ & $ 34$  & $ 26$ & $ 20$ & $ 14 $ & $ 10 $  & $  6 $ & $  4 $ & $ 2  $ & $ 1  $  & $ 0  $ & $    $  & $    $ & $4752 $ & $20592   $  \\
    $17$ & $ 25$ & $ 45$ & $ 56$ & $ 64$  & $ 65$ & $ 64$ & $ 58$ & $ 52$  & $ 43$ & $ 35$ & $ 26 $ & $ 20 $  & $ 14 $ & $ 10 $ & $ 6  $ & $ 4  $  & $ 2  $ & $ 1  $  & $    $ & $7722 $ & $28314   $  \\
    $18$ & $ 25$ & $ 45$ & $ 56$ & $ 64$  & $ 65$ & $ 64$ & $ 58$ & $ 52$  & $ 43$ & $ 35$ & $ 26 $ & $ 20 $  & $ 14 $ & $ 10 $ & $ 6  $ & $ 4  $  & $ 2  $ & $ 1  $  & $ 0  $ & $7722 $ & $36036   $  \\
 \hline\hline
    \end{tabular}
 \caption{Number of SO(3) multiplets with $n_F=1$ and fixed $j$ and $n_B$. $N_s$ is the number of basis vectors
   with given number of bosonic quanta, $n_B$, while $\Sigma$ gives the cumulative size up to $n_B$.}
 \label{tableF1}
 \end{table}
\end{landscape}

For $j=3/2$ new fermionic creators can be constructed beginning with $n_B=3$. In this case there are
two old creators $(fa)_{3/2} \Tr(A)$ and $(fa.A)_{3/2}$, corresponding to coupling $3/2\otimes 0\rightarrow 3/2$,
and $(1/2\otimes 1) \otimes 2\rightarrow 3/2$
\footnote{The trace in $A$ is linearly dependent with the the first creator
and there is no gauge invariant, j=1, combination of $a$'s}.
However there are {\em two} ways to realize the last coupling. Hence there must exist an independent creator
$(fa.\tilde{A})_{3/2}$, where $\tilde{A}^{i\sigma}$ denotes generically the bilinear form of bosonic creators
convoluted with all appropriate Clebsch-Gordan coefficients. Now, to generate all independent states
with $n_F=1,j=3/2$ and $n_B$ bosons we have to act with $(fa)_{3/2}$ on the $n_F=0,j=0,n_B-1$ basis,
with $(fa.A)_{3/2}$ and $(fa.\tilde{A})_{3/2}$ on the $n_F=0,j=0, n_B-3$ basis, with $(fa.A^2)_{3/2}$
and $(fa.A.\tilde{A})_{3/2}$
on the $n_F=0,j=0, n_B-5$ basis, and finally with $(fa.A.A.\tilde{A})_{3/2}$ on the $n_F=0,j=0, n_B-7$ basis.
Therefore
\eq
{\cal N}(1,1/2,n_B)={\cal N}(0,0,n_B-1)+2({\cal N}(0,0,n_B-3)+{\cal N}(0,0,n_B-5))+{\cal N}(0,0,n_B-7).
\eqx
This explains the second column of Table \ref{tableF1}.

\noindent {\bf D.4} $n_F=2, j=0.$\newline
 A general two-fermion creator can have three covariant forms: (a) an SO(3) scalar
 $F=f_b^{\dagger 1/2}f_b^{\dagger -1/2}$ , (b) an SO(3) scalar symmetric in color indices
 $^SF_{bc}=f_b^{\dagger 1/2}f_c^{\dagger -1/2}-f_b^{\dagger -1/2}f_c^{\dagger 1/2} $,
 and (c) an SO(3) vector antisymmetric in color indices
$^AF_{bc}^i$ corresponding to a coupling: $
1/2\otimes1/2\rightarrow 1$. To construct gauge invariant creators
from (2) define the matrix $(^SFA)^{ik}=^SF_{bc}a_b^{\dagger i}
a_c^{\dagger k}$. Independent symmetric creators are \eq
 \Tr(^SFA),\; \Tr(^SFA.A),\; \Tr(^SFA.A^2)\; and\; F.
\eqx
 Antisymmetric creators are constructed from the matrix
 $(^AFA)^{il}=^AF_{bc}^i a_b^{\dagger j} a_c^{\dagger k}\epsilon^{jkl}$. Cayley-Hamilton
 equation allows to construct two independent creators in this case
 \eq
 \Tr(^AFA) ,\;\; \Tr(^AFA.A).
 \eqx
Then the $n_F=2,j=0$,basis with $n_B$ quanta can be obtained by
acting with all these creators on the appropriate $j=0$ bases with
lower $n_B$ \footnote{This decomposition is analogous to that of
van Baal \cite{vBN}.}. As a consequence
\eq {\cal
N}(2,0,n_B)={\cal N}(0,0,n_B)+2({\cal N}(0,0,n_B-2)+{\cal
N}(0,0,n_B-4))+{\cal N}(0,0,n_B-6).
\eqx
Which is readily
satisfied by the first column of Table \ref{tableF2}.

\noindent {\bf D.5} $ n_F=3, j=1/2. $\newline It is left as an
exercise for the reader to prove why the first column of Table
\ref{tableF3} satisfies the following recursion for odd $n_B$ \eq
{\cal N}(3,1/2,n_B)={\cal N}(0,0,n_B-1)+2{\cal N}(0,0,n_B-3)+ 4
{\cal N}(0,0,n_B-5)+ 3 {\cal N}(0,0,n_B-7). \eqx

  \begin{landscape}
   \begin{table}[tbp]
    \begin{tabular}{cccccccccccccccccccccc} \hline\hline
    $J$ & $ 0 $ & $ 1 $ & $ 2 $ & $ 3 $  & $ 4 $ & $ 5 $ & $ 6 $ & $ 7 $  & $ 8 $ & $ 9 $ & $ 10 $ & $ 11 $  & $ 12 $ & $ 13 $ & $ 14 $ & $ 15 $  & $ 16 $ & $ 17 $  & $ 18 $ & $ N_s $ & $ \Sigma $  \\
    $n_B$ & $   $ & $   $ & $   $ & $   $  & $   $ & $   $ & $   $ & $   $  & $   $ & $   $ & $    $ & $    $  & $    $ & $    $ & $    $ & $    $  & $    $ & $    $  & $    $ & $     $ & $        $  \\
    \hline
    $0$  & $ 1 $ & $ 0 $ & $   $ & $   $  & $   $ & $   $ & $   $ & $   $  & $   $ & $   $ & $    $ & $    $  & $    $ & $    $ & $    $ & $    $  & $    $ & $    $  & $    $ & $  1  $ & $    1   $  \\
    $1$  & $ 1 $ & $ 1 $ & $ 1 $ & $   $  & $   $ & $   $ & $   $ & $   $  & $   $ & $   $ & $    $ & $    $  & $    $ & $    $ & $    $ & $    $  & $    $ & $    $  & $    $ & $  9  $ & $   10   $  \\
    $2$  & $ 3 $ & $ 1 $ & $ 3 $ & $ 0 $  & $   $ & $   $ & $   $ & $   $  & $   $ & $   $ & $    $ & $    $  & $    $ & $    $ & $    $ & $    $  & $    $ & $    $  & $    $ & $ 21  $ & $   31   $  \\
    $3$  & $ 3 $ & $ 4 $ & $ 5 $ & $ 2 $  & $ 1 $ & $   $ & $   $ & $   $  & $   $ & $   $ & $    $ & $    $  & $    $ & $    $ & $    $ & $    $  & $    $ & $    $  & $    $ & $ 63  $ & $   94   $  \\
    $4$  & $ 6 $ & $ 4 $ & $ 9 $ & $ 3 $  & $ 3 $ & $ 0 $ & $   $ & $   $  & $   $ & $   $ & $    $ & $    $  & $    $ & $    $ & $    $ & $    $  & $    $ & $    $  & $    $ & $111  $ & $  205   $  \\
    $5$  & $ 6 $ & $ 8 $ & $13 $ & $ 8 $  & $ 6 $ & $ 2 $ & $ 1 $ & $   $  & $   $ & $   $ & $    $ & $    $  & $    $ & $    $ & $    $ & $    $  & $    $ & $    $  & $    $ & $240  $ & $  445   $  \\
    $6$  & $10 $ & $ 8 $ & $19 $ & $10 $  & $11 $ & $ 3 $ & $ 3 $ & $ 0 $  & $   $ & $   $ & $    $ & $    $  & $    $ & $    $ & $    $ & $    $  & $    $ & $    $  & $    $ & $370  $ & $  815   $  \\
    $7$  & $10 $ & $14 $ & $24 $ & $18 $  & $17 $ & $ 9 $ & $ 6 $ & $ 2 $  & $ 1 $ & $   $ & $    $ & $    $  & $    $ & $    $ & $    $ & $    $  & $    $ & $    $  & $    $ & $675  $ & $ 1490   $  \\
    $8$  & $15 $ & $14 $ & $32 $ & $21 $  & $25 $ & $12 $ & $11 $ & $ 3 $  & $ 3 $ & $ 0 $ & $    $ & $    $  & $    $ & $    $ & $    $ & $    $  & $    $ & $    $  & $    $ & $960  $ & $ 2450   $  \\
    $9$  & $15 $ & $21 $ & $39 $ & $32 $  & $34 $ & $22 $ & $18 $ & $ 9 $  & $ 6 $ & $ 2 $ & $ 1  $ & $    $  & $    $ & $    $ & $    $ & $    $  & $    $ & $    $  & $    $ & $1575 $ & $ 4025   $  \\
    $10$ & $21 $ & $21 $ & $49 $ & $36 $  & $45 $ & $27 $ & $27 $ & $12 $  & $11 $ & $ 3 $ & $ 3  $ & $ 0  $  & $    $ & $    $ & $    $ & $    $  & $    $ & $    $  & $    $ & $2121 $ & $ 6146   $  \\
    $11$ & $21 $ & $30 $ & $57 $ & $50 $  & $57 $ & $42 $ & $38 $ & $23 $  & $18 $ & $ 9 $ & $ 6  $ & $ 2  $  & $  1 $ & $    $ & $    $ & $    $  & $    $ & $    $  & $    $ & $3234 $ & $ 9380   $  \\
    $12$ & $28 $ & $30 $ & $69 $ & $55 $  & $71 $ & $49 $ & $51 $ & $29 $  & $27 $ & $12 $ & $11  $ & $ 3  $  & $  3 $ & $  0 $ & $    $ & $    $  & $    $ & $    $  & $    $ & $4186 $ & $13566   $  \\
    $13$ & $28 $ & $40 $ & $79 $ & $72 $  & $86 $ & $68 $ & $67 $ & $46 $  & $39 $ & $23 $ & $18  $ & $ 9  $  & $  6 $ & $  2 $ & $ 1  $ & $    $  & $    $ & $    $  & $    $ & $6048 $ & $19614   $  \\
    $14$ & $36 $ & $40 $ & $93 $ & $78 $  & $103$ & $77 $ & $84 $ & $55 $  & $53 $ & $29 $ & $27  $ & $12  $  & $ 11 $ & $  3 $ & $ 3  $ & $ 0  $  & $    $ & $    $  & $    $ & $7596 $ & $27210   $  \\
    $15$ & $36 $ & $52 $ & $104$ & $98 $  & $121$ & $101$ & $104$ & $78 $  & $71 $ & $47 $ & $39  $ & $23  $  & $ 18 $ & $  9 $ & $ 6  $ & $ 2  $  & $ 1  $ & $    $  & $    $ & $10530$ & $37740   $  \\
    $16$ & $45 $ & $52 $ & $120$ & $105$  & $141$ & $112$ & $125$ & $90 $  & $90 $ & $57 $ & $53  $ & $29  $  & $ 27 $ & $ 12 $ & $11  $ & $ 3  $  & $ 3  $ & $  0 $  & $    $ & $12915$ & $50655   $  \\
 \hline\hline
    \end{tabular}
 \caption{Number of SO(3) multiplets with $n_F=2$ and fixed $j$ and $n_B$. $N_s$ is the number of basis vectors
   with given number of bosonic quanta, $n_B$, while $\Sigma$ gives the cumulative size up to $n_B$.}
 \label{tableF2}
 \end{table}
\end{landscape}

  \begin{landscape}
   \begin{table}[tbp]
    \begin{tabular}{ccccccccccccccccccccrr} \hline\hline
    $J$ & $ {1\over 2} $ & ${3\over 2} $ & ${5\over 2} $ & ${7\over 2}$ & ${9\over 2}$ & ${11\over 2}$ & $ {13\over 2}$ & $ {15\over 2}$  & $ {17\over 2}$ & $ {19\over 2}$ & $ {21\over 2}$ & ${23\over 2}$  & ${25\over 2}$ & ${27\over 2}$ & ${29\over 2}$ & $ {31\over 2}$  & $ {33\over 2}$ & $ {35\over 2}$  & $ {37\over 2}$ & $ N_s $ & $ \Sigma $  \\
    $n_B$ & $   $ & $   $ & $   $ & $   $  & $   $ & $   $ & $   $ & $   $  & $   $ & $   $ & $    $ & $    $  & $    $ & $    $ & $    $ & $    $  & $    $ & $    $  & $    $ & $     $ & $        $  \\
    \hline
    $0$  & $ 0 $ & $ 1 $ & $   $ & $   $  & $   $ & $   $ & $   $ & $   $  & $   $ & $   $ & $    $ & $    $  & $    $ & $    $ & $    $ & $    $  & $    $ & $    $  & $    $ & $    4$ & $    4   $  \\
    $1$  & $ 1 $ & $ 1 $ & $ 0 $ & $   $  & $   $ & $   $ & $   $ & $   $  & $   $ & $   $ & $    $ & $    $  & $    $ & $    $ & $    $ & $    $  & $    $ & $    $  & $    $ & $    6$ & $   10   $  \\
    $2$  & $ 3 $ & $ 4 $ & $ 2 $ & $ 1 $  & $   $ & $   $ & $   $ & $   $  & $   $ & $   $ & $    $ & $    $  & $    $ & $    $ & $    $ & $    $  & $    $ & $    $  & $    $ & $   42$ & $   52   $  \\
    $3$  & $ 3 $ & $ 6 $ & $ 3 $ & $ 1 $  & $ 0 $ & $   $ & $   $ & $   $  & $   $ & $   $ & $    $ & $    $  & $    $ & $    $ & $    $ & $    $  & $    $ & $    $  & $    $ & $   56$ & $  108   $  \\
    $4$  & $ 7 $ & $11 $ & $ 9 $ & $ 6 $  & $ 2 $ & $ 1 $ & $   $ & $   $  & $   $ & $   $ & $    $ & $    $  & $    $ & $    $ & $    $ & $    $  & $    $ & $    $  & $    $ & $  192$ & $  300   $  \\
    $5$  & $ 8 $ & $13 $ & $11 $ & $ 8 $  & $ 3 $ & $ 1 $ & $ 0 $ & $   $  & $   $ & $   $ & $    $ & $    $  & $    $ & $    $ & $    $ & $    $  & $    $ & $    $  & $    $ & $  240$ & $  540   $  \\
    $6$  & $12 $ & $22 $ & $21 $ & $17 $  & $11 $ & $ 6 $ & $ 2 $ & $ 1 $  & $   $ & $   $ & $    $ & $    $  & $    $ & $    $ & $    $ & $    $  & $    $ & $    $  & $    $ & $  600$ & $ 1140   $  \\
    $7$  & $14 $ & $24 $ & $24 $ & $21 $  & $13 $ & $ 8 $ & $ 3 $ & $ 1 $  & $ 0 $ & $   $ & $    $ & $    $  & $    $ & $    $ & $    $ & $    $  & $    $ & $    $  & $    $ & $  720$ & $ 1860   $  \\
    $8$  & $20 $ & $35 $ & $38 $ & $36 $  & $27 $ & $19 $ & $11 $ & $ 6 $  & $ 2 $ & $ 1 $ & $    $ & $    $  & $    $ & $    $ & $    $ & $    $  & $    $ & $    $  & $    $ & $ 1500$ & $ 3360   $  \\
    $9$  & $21 $ & $39 $ & $42 $ & $40 $  & $32 $ & $23 $ & $13 $ & $ 8 $  & $ 3 $ & $ 1 $ & $ 0  $ & $    $  & $    $ & $    $ & $    $ & $    $  & $    $ & $    $  & $    $ & $ 1750$ & $ 5110   $  \\
    $10$ & $29 $ & $52 $ & $60 $ & $61 $  & $52 $ & $42 $ & $29 $ & $19 $  & $11 $ & $ 6 $ & $ 2  $ & $ 1  $  & $    $ & $    $ & $    $ & $    $  & $    $ & $    $  & $    $ & $ 3234$ & $ 8344   $  \\
    $11$ & $31 $ & $56 $ & $65 $ & $67 $  & $58 $ & $48 $ & $34 $ & $23 $  & $13 $ & $ 8 $ & $ 3  $ & $ 1  $  & $  0 $ & $    $ & $    $ & $    $  & $    $ & $    $  & $    $ & $ 3696$ & $12040   $  \\
    $12$ & $39 $ & $73 $ & $87 $ & $92 $  & $86 $ & $75 $ & $58 $ & $44 $  & $29 $ & $19 $ & $11  $ & $ 6  $  & $  2 $ & $  1 $ & $    $ & $    $  & $    $ & $    $  & $    $ & $ 6272$ & $18312   $  \\
    $13$ & $42 $ & $77 $ & $93 $ & $100$  & $93 $ & $83 $ & $66 $ & $50 $  & $34 $ & $23 $ & $13  $ & $ 8  $  & $  3 $ & $  1 $ & $ 0  $ & $    $  & $    $ & $    $  & $    $ & $ 7056$ & $25368   $  \\
    $14$ & $52 $ & $96 $ & $119$ & $131$  & $127$ & $118$ & $100$ & $81 $  & $60 $ & $44 $ & $29  $ & $19  $  & $ 11 $ & $  6 $ & $ 2  $ & $ 1  $  & $    $ & $    $  & $    $ & $11232$ & $36600   $  \\
    $15$ & $54 $ & $102$ & $126$ & $139$  & $137$ & $128$ & $109$ & $91 $  & $68 $ & $50 $ & $34  $ & $23  $  & $ 13 $ & $  8 $ & $ 3  $ & $ 1  $  & $ 0  $ & $    $  & $    $ & $12480$ & $49080   $  \\
    $16$ & $66 $ & $123$ & $156$ & $176$  & $177$ & $171$ & $153$ & $132$  & $106$ & $83 $ & $60  $ & $44  $  & $ 29 $ & $ 19 $ & $ 11 $ & $ 6  $  & $ 2  $ & $ 1  $  & $    $ & $18900$ & $67980   $  \\
 \hline\hline
    \end{tabular}
 \caption{Number of SO(3) multiplets with $n_F=3$ and fixed $j$ and $n_B$. $N_s$ is the number of basis vectors in all angular momentum channels
   with given number of bosonic quanta, $n_B$, while $\Sigma$ gives the cumulative size up to $n_B$.}
 \label{tableF3}
 \end{table}
 \end{landscape}

\end{document}